\documentclass[12pt]{iopart}
\usepackage{iopams}  
\usepackage{graphicx}
\usepackage{epstopdf}
\usepackage{subfig}

\newcommand{\tk}{T_{\mathrm{kin}}}
\newcommand{\tch}{T_{\mathrm{ch}}}
\begin{document}

\title[]{Kinetic freeze-out in central heavy-ion collisions  between 7.7 and  2760~GeV per nucleon pair}
\author{Ivan Melo$^{1}$ and Boris Tom\'a\v{s}ik$^{2,3}$}
\address{%
$^1$ \v{Z}ilinsk\'a Univerzita, Akademick\'a 1, 01026 \v{Z}ilina, Slovakia}
\address{%
$^2$ Univerzita Mateja Bela, Tajovsk\'eho 40, 97401 Bansk\'a Bystrica, Slovakia
}
\address{%
$^3$ Czech Technical University in Prague, FNSPE, B\v{r}ehov\'a 7, 11519 Prague, 
Czechia
}

\date{\today}


\begin{abstract}
We fit the single-particle $p_t$ spectra of identified pions, kaons, and (anti)protons from 
central collisions of gold or lead nuclei at energies between 7.7 and 2760~GeV per nucleon pair. 
Blast wave model with included resonance production and with an assumption of partial 
chemical equilibrium is used and the fits are performed with the help of a Gaussian emulator process. 
A kinetic freeze-out temperature is found about 100~MeV for the lowest collision energies and 80~MeV
at the LHC. The average transverse expansion velocity grows with increasing $\sqrt{s_{NN}}$
from 0.45 to 0.65. Due to partial chemical equilibrium, the influence of resonance decays
on the shape of the $p_t$ spectra for $\sqrt{s_{NN}}$ above 27~GeV is small.
\end{abstract}



\section{Introduction}
\label{s:intro}

Phase diagram of strongly interacting matter can be explored with the help of heavy-ion collisions at various 
collision energies. The higher the collision energy, the closer is the created hot matter to baryon-antibaryon 
symmetry. It is the purpose of the RHIC Beam Energy Scan (BES) programme to study the dependences of various 
observables on the collision energy. The ultimate goal of the programme is to explore the QCD phase diagram and
possibly find the critical point of the deconfinement phase transition. 

Bulk evolution of the fireball is reflected in the distributions of hadrons which are emitted at the point of their kinetic
freeze-out. This is the last moment of the lifetime of the fireball, after it has expanded and cooled down so that 
no hadrons scatter anymore. By combining information from single-particle spectra of different hadron species it is 
possible to reconstruct the freeze-out state of the fireball. This mainly means revealing its temperature and the transverse 
expansion velocity. This state is the result of the previous expansion due to initial conditions and internal pressure and that, 
in turn, is conditioned by the properties of the matter: its Equation of State and the transport coefficients. 
Thus---in principle---the knowledge of the final state allows to deduce the previous evolution and the properties of the 
matter.  

It is then interesting to study the freeze-out state for different collision energies and observe how it changes when 
the energy is varied. 

In this paper we fit the single-particle spectra of pions, kaons, and (anti)protons from central Au+Au or Pb+Pb collisions
in the energy range $\sqrt{s_{NN}} = 7.7$ to  2760~GeV. We use the blast wave model. 
Also included  is the production of resonances from the blast-wave model, which then decay and contribute to the numbers of 
stable hadrons.  
The novelty of our approach is in consistent treatment of the fireball at lower temperature. Fits to the abundances of the 
identified hadrons indicate chemical freeze-out temperatures above 145~MeV \cite{STAR_7-39,STAR_62-200,Milano:2013sza}. 
As we will see, our results for the \emph{kinetic} freeze-out temperature are at least by 40~MeV lower. Despite the lower 
temperature, the abundances must be observed. This implies chemical potentials which are individual for every hadron species. 
In a Fermi-Dirac or Bose-Einstein distribution, the presence of chemical potential may change the momentum distribution. 
Therefore, they are consistently calculated and used here. In practice, as we will show in the paper, at the lower temperature
the dominant production moves away from the decays of resonances to a larger portion of hadrons being produced thermally. 
%

The inclusion of resonances is computationally very involved\footnote{%
 Note, however, that in \cite{Mazeliauskas:2018irt} a new computational approach was developed which makes the 
 calculation of the spectrum with included resonance production more effective.}. 
Therefore, we use DRAGON  \cite{DRAGON,Tomasik:2016skq}---a Monte Carlo generator of the final state 
hadrons---in order to find the best theoretical model. 
The Bayesian fits then use Gaussian process emulator
for the Markov Chain MC procedure of looking for the best model parameters \cite{madai4}. 

There are other similar analyses published in the literature. 
First of all, the spectra are usually fitted  by the experimental collaborations which measure them. ALICE has measured and
fitted their $p_t$ spectra from Pb+Pb collisions at $\sqrt{s_{NN}} = 2.76$~GeV in their original publication \cite{ALICE_piKp}.
In these fits, however, only directly produced hadrons are taken into account and no resonance decays. The same is true
for Au+Au collisions at all lower energies studied here, which were measured and fitted by the STAR collaboration
\cite{STAR_7-39,STAR_62-200}. We have fitted the spectra from ALICE in \cite{Melo:2015wpa} with the model used here, 
but without the assumption of partial chemical equilibrium. These spectra have also been fitted in \cite{Begun:2013nga,Begun:2014rsa} 
with the help of the Cracow single freeze-out model \cite{Broniowski:2001we,Chatterjee:2014lfa} 
with chemical non-equilibrium. The single freeze-out model has been 
augmented with sample averaging over events with varying temperature in \cite{Prorok:2015vxa,Prorok:2018okq}
and fitted to the same data. The blast wave model has been fitted to single-particle distributions from nuclear collisions in a wide 
interval of collision energies in \cite{Rode:2018hlj}. There are slight differences to our present treatment, however: 
that source is cut-off in space-time rapidity, and the composition of resonance contributions is taken according to
chemical equilibrium.   In \cite{Li:2018jnm} the spectra are fitted with a two component blast-wave model, which, however, 
misses the resonance contribution completely.

Just a few days before finishing this manuscript a similar paper appeared, in which the spectra from ALICE are fitted with the
blast-wave model \cite{Mazeliauskas:2019ifr}. Resonance decays are included there with the help of a new treatment 
introduced in \cite{Mazeliauskas:2018irt}.

Our results show that the fireball freezes out at lower temperature and stronger transverse expansion if the collision
energy is increased. In addition to that, we also show that for higher collision energies, thanks to the low temperature and the partial chemical 
equilibrium, the spectra with full resonance production are very similar to those calculated for only directly produced 
particles. 

This paper is structured as follows. In the next Section we explain the model. Section \ref{s:dam} deals with the introduction 
of data which are analysed. In Section \ref{s:res} we present our main results on the temperature and transverse flow
and in Section~\ref{s:anat} we discuss in detail the individual contributions to the spectra from the decays of 
different resonances. Some semi-quantitative estimates of the $p_t$ dependence of these contributions are 
deferred to \ref{s:hadmom}. We conclude in Section~\ref{s:conc}.


\section{The model}
\label{s:model}

The theoretical model used in our analysis is based on the well-known blast wave model
\cite{Siemens:1978pb,Schnedermann:1993ws,Csorgo:1995bi,Tomasik:1999cq,Retiere:2003kf}. 
The model 
is formulated with the help of its emission function, which is the Wigner function. For hadrons of the type 
$i$ it reads
\begin{eqnarray}
\nonumber
S(x,p)\, d^4x & = & g_i \,  \frac{m_t\, \cosh(\eta-y)}{(2\pi)^3}  \left ( \exp\left ( \frac{p_\mu u^\mu - \mu_i }{\tk} \right ) + s_i \right )^{-1}
\theta\left ( 1 - \frac{r}{R} \right )
\nonumber \\  && \qquad \qquad 
\times r\, dr\, d\varphi\, \delta(\tau - \tau_0)\, 
\tau\, d\tau\, d\eta\,  . 
\label{e:Sfun}
\end{eqnarray}
Usually, longitudinal flow dominates the fireball expansion in ultrarelativistic heavy-ion collisions. Therefore, 
one uses longitudinal 
proper time $\tau = \sqrt{t^2 - z^2}$ and space-time rapidity 
$\eta = \frac{1}{2}\ln((t+z)/(t-z))$. In the transverse plane, polar coordinates $r$, $\varphi$ are used. 
Furthermore, 
$\tk$ is the kinetic freeze-out temperature, $m_t$ the transverse mass and $\mu_i$ the chemical potential.
We use the proper quantum statistical distributions with $s_i = 1$ 
$(-1)$ for fermions (bosons) and $g_i$ is the spin degeneracy. Every isospin state 
is treated separately. The freeze-out time does not depend on radial coordinate, 
but it does depend on the longitudinal coordinate implicitly via 
$\tau = \tau_0$, i.e. $t = \sqrt{\tau_0^2 + z^2}$. The density is distributed uniformly within the radius $R$. 

The expansion of 
the fireball is represented by the velocity field
\begin{equation}
u^\mu  = \left (  \cosh\eta_t \cosh\eta,\, \sinh\eta_t\cos\varphi, 
 \sinh\eta_t \sin\varphi,\, \cosh\eta_t\sinh\eta\right )
\end{equation}
where the transverse velocity is such that 
\begin{equation}
v_t = \tanh\eta_t = \eta_f \left ( \frac{r}{R} \right )^n\,  .
\label{e:vn}
\end{equation}
In this relation $\eta_f$ parametrises the transverse flow gradient and $n$ the profile 
of the transverse velocity. The mean transverse velocity is then
\begin{equation}
\label{e:meanv}
\langle v_t \rangle = \frac{2}{n+2} \eta_f\,  .
\end{equation}
This parametrisation of the transverse velocity is taken to be the same as in \cite{STAR_7-39,STAR_62-200,Mazeliauskas:2018irt,ALICE_piKp}.

The transverse size $R$ and the freeze-out proper time $\tau_0$ influence the total 
normalisations of the transverse momentum spectra. 
These parameters also influence the sizes of the HBT radii. A fit which obtains $R$ and $\tau_0$ should thus also take into account the data on HBT radii. Due to computational complexity of such a problem we choose not to embark on this way and remain without the sensitivity to these geometrical parameters.

From the emission function, one obtains the single-particle spectrum of directly produced hadrons as
\begin{equation}
\label{e:sspec}
E\frac{d^3N}{dp^3} = \int_\Sigma S(x,p)\, d^4x\,  ,
\end{equation}
where the integration runs over the whole freeze-out hypersurface. If one replaces 
the quantum-statistical distribution in eq.~(\ref{e:Sfun}) by the classical Boltzmann
distribution and performs some of the integrations in eq.~(\ref{e:sspec}), one arrives at 
\begin{equation}    
E\frac{d^3N}{dp^3} = \frac{m_t \tau_0}{2\pi^2} e^{\mu_i/\tk}  
\int_0^R r\, I_0\left ( \frac{p_t}{\tk}\sinh\eta_t(r) \right ) \,
 K_1\left ( \frac{m_t}{\tk}\cosh\eta_t(r) \right )\, dr\,  .
\label{e:dirspec}
\end{equation}
This formula is rather easy to evaluate and thus it is often used in the spectra fitting. 

Resonances are emitted as described by the emission function in eq.~(\ref{e:Sfun}) and 
then decay exponentially in time with the mean lifetime given by the inverse of the resonance width. 
All matrix elements for 
the decays are assumed to be constant and thus the decay is determined by the 
phase-space only. We include baryon resonances up to the mass of 2~GeV and 
mesonic resonances up to 1.5~GeV. 
An analytical expression for the calculation of the spectra from resonance decays 
has been derived \cite{Schnedermann:1993ws}, but it is too cumbersome for frequent evaluation 
within a fitting routine\footnote{%
Note, however, that after we finished our fits, a new treatment has been proposed \cite{Mazeliauskas:2018irt}
which should allow 
for the evaluation of spectra including resonance decays without the need of Monte Carlo simulations. 
}.
Therefore, we use DRAGON for theoretical simulation of the single-particle spectra \cite{DRAGON,Tomasik:2016skq}. 
This is a Monte Carlo event generator, which produces hadrons (including resonances) according to the emission 
function~(\ref{e:Sfun}). 

For the contribution from the resonance decays it is also necessary to specify the abundances of every individual 
resonance species. The final state composition is measured and it is usually found to be in good agreement 
with the Statistical Hadronisation Model (SHM), which assumes chemical equilibrium with chemical freeze-out temperature
$\tch$. However, our results will indicate that the fireball freezes out kinetically at $\tk$ much lower than $\tch$. 
After the chemical freeze-out, inelastic collisions become rare\footnote{%
The relevant (inverse) time scale is the expansion rate ($\partial^\mu u_\mu$), so ``rare'' means a reaction rate much lower than this.
}%
and the system slips out of the full chemical equilibrium. Nevertheless, in order to keep  the abundances of final state stable species  fixed, 
each species develops 
a nonzero temperature-dependent chemical potential $\mu_i$.
At this phase interactions still maintain the {\em partial chemical equilibrium} between the lowest state stable hadrons and the resonances
through which they interact \cite{Bebie:1991ij}. Also, elastic collisions keep the local thermal equilibrium until the fireball 
freezes out completely.

Indeed, not all inelastic collisions drop out below $\tch$. 
For example, $\rho \leftrightarrow \pi \pi$ process remains fast enough to regenerate $\rho$ resonances 
and, as a consequence, neither the pion number $N_{\pi}$, nor the $\rho$ number $N_{\rho}$ is fixed, 
but rather their combination $N_{\pi} + 2 N_{\rho}$. 
For the chemical potential this implies $\mu_\rho = 2\mu_\pi$.
The number $N_{\rho}$ continually drops with the temperature decrease 
while $N_{\pi}$ increases 
until they are finally fixed at $\tk$. 
Another example may be the $\Delta^{++}$ resonance which appears in the process $\pi^+ p \leftrightarrow \Delta^{++}$. 
Hence, $\mu_{\Delta^{++}} = \mu_\pi + \mu_p$. In our calculation, 254 resonance species are included in this way. They are in partial
equilibrium with their decay products. Their chemical potentials 
$\mu_R$ are given by the sum of those of the stable decay 
products $\mu_i$ multiplied by the effective numbers of stable hadrons $i$ produced on average from a decay of a resonance $R$ \cite{Bebie:1991ij}
\begin{equation}
\mu_R = \sum_i N_{i,R}\, \mu_i\,  .
\end{equation}
Resonances are then let to decay so that in the end we look only at the stable hadrons. 
In our calculation, the stable species with their own chemical potentials are:
$\pi^+$, $\pi^-$, $\pi^0$, $K^+$, $K^-$, $K^0$, $\bar K^0$, $p$, $n$, $\bar p$, $\bar n$, $\Lambda$, $\bar \Lambda$, $\Sigma^+$,
$\Sigma^-$, $\bar \Sigma^+$, $\bar \Sigma^-$, $\Xi^0$, $\Xi^-$, $\bar \Xi^0$, $\bar \Xi^-$, $\Omega$, $\bar \Omega$.

\begin{figure}
        \begin{center}
	\includegraphics[width=0.45\textwidth]{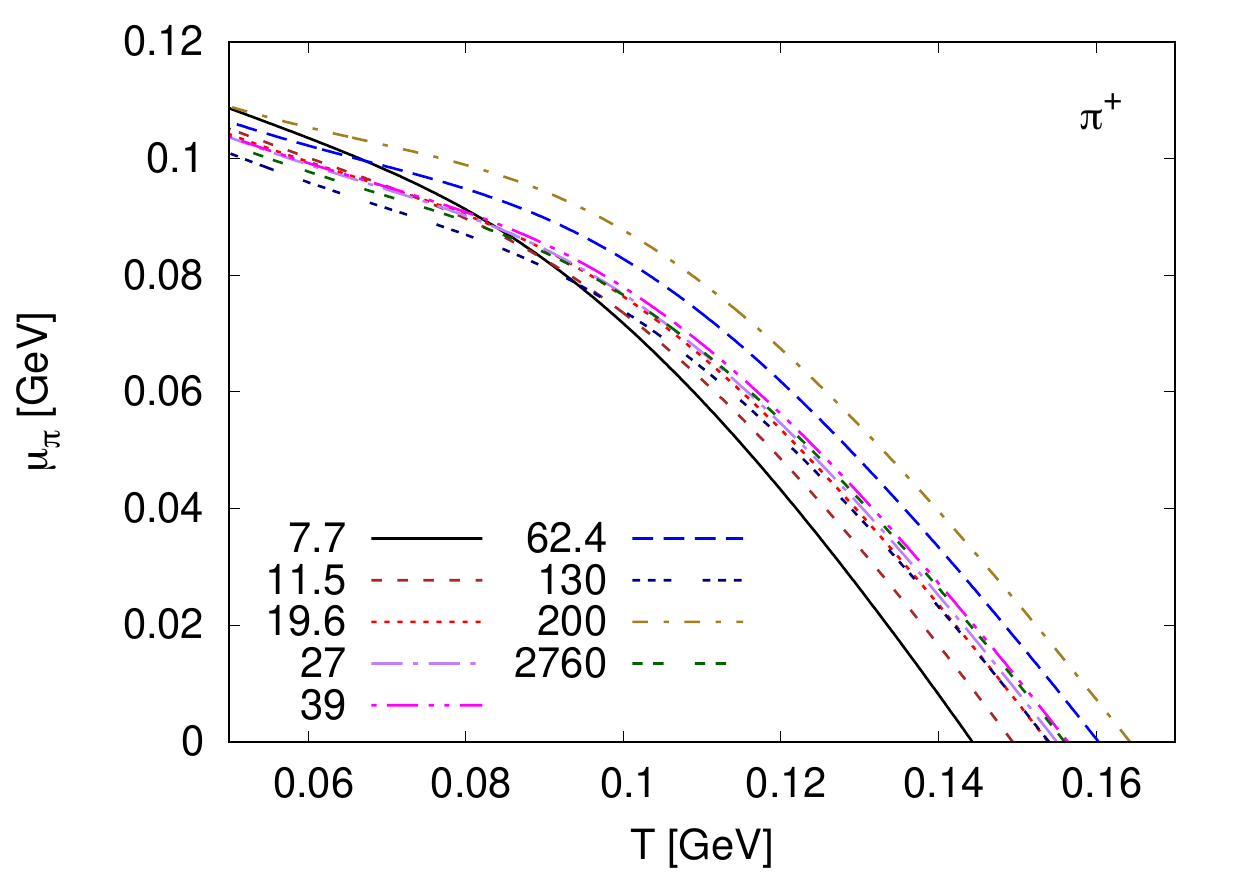}
	\includegraphics[width=0.45\textwidth]{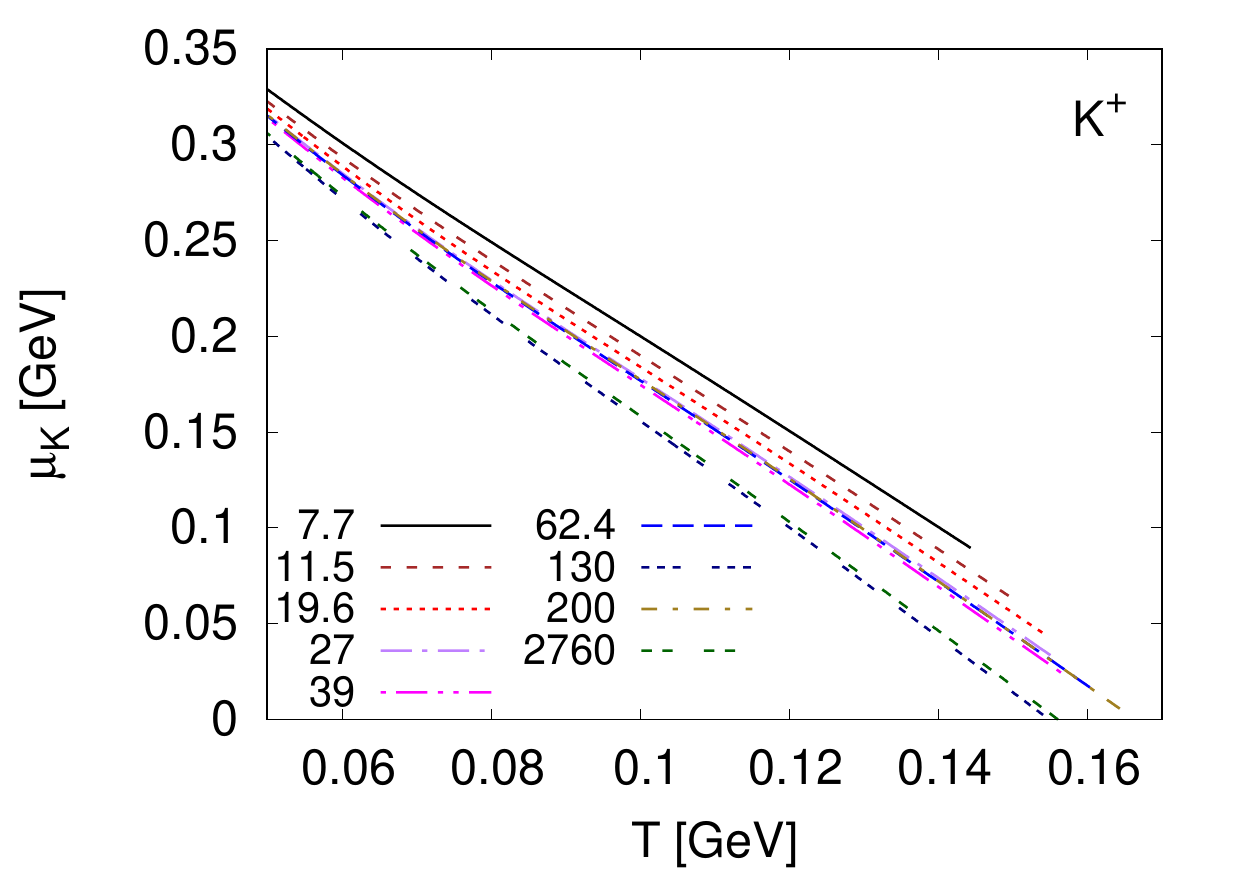}
	\includegraphics[width=0.45\textwidth]{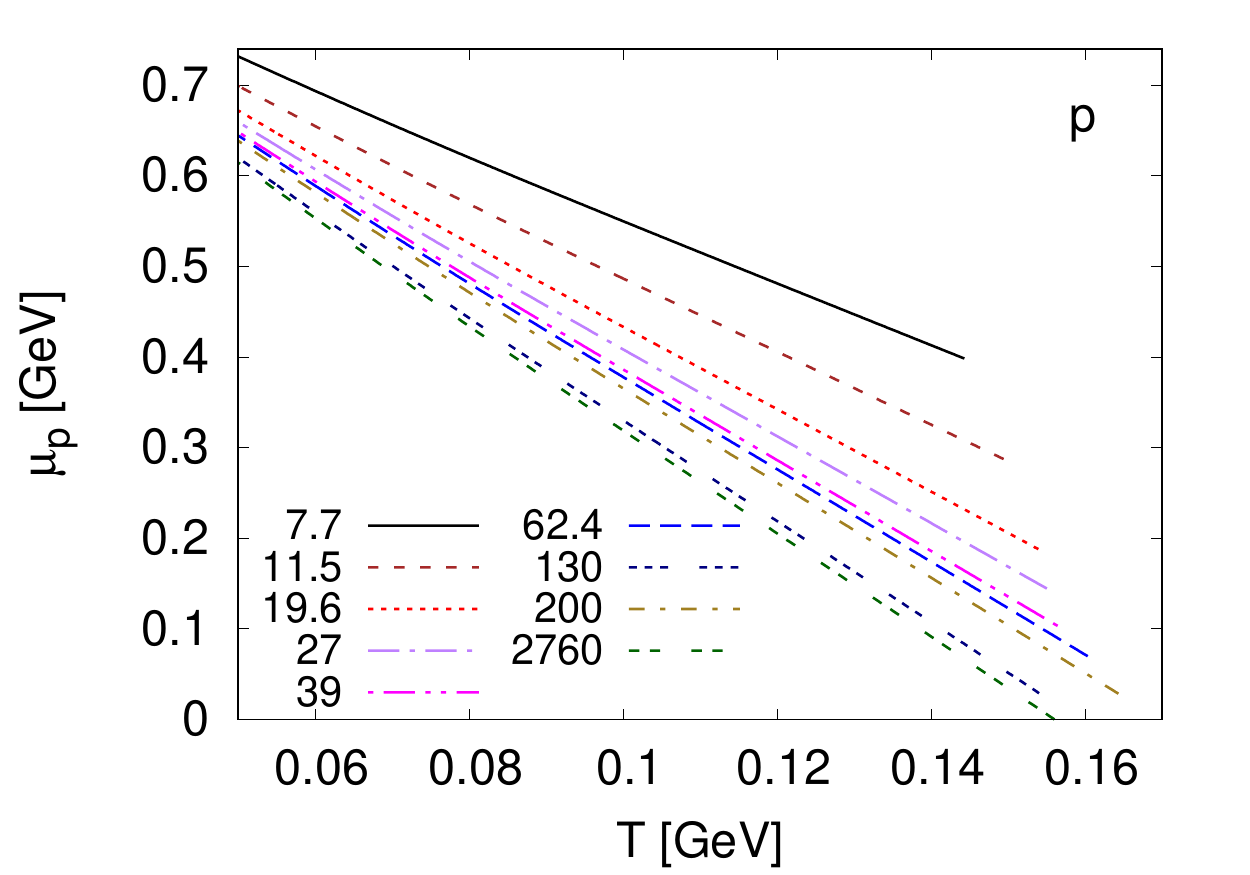}
	\end{center}
\caption{Temperature dependence of the  chemical potentials for $\pi^+$, $K^+$, and protons.}
\label{ChemPotentials} 
\end{figure}

As an example,  temperature dependences of the chemical potentials for positive pions, kaons, and protons are shown in Fig.~\ref{ChemPotentials} 
for different collision energies from the RHIC BES program and the LHC. 
They were evolved from the temperatures of chemical freeze-out towards lower temperatures.
As argued in \cite{Bebie:1991ij}, the evolutions are formulated by requiring that the ratios $n_i/s$ stay independent of temperature. 
Here, $s$ is the entropy density and $n_i$ the total effective density of stable hadron species $i$, i.e.~ the density which includes hadrons
which will be produced when all resonances decay. 
The evolutions always start at the highest temperature in the state of chemical equilibrium. Thus, e.g., protons start with $\mu_p = \mu_B$ 
at $\tch$, positive kaons start with $\mu_K = \mu_S$, and pions start with $\mu_\pi = 0$ (since we neglect the isospin chemical potential). 
The initial values for $\tch$, $\mu_B$, and $\mu_S$ for each collision energy have been taken from \cite{STAR_7-39,STAR_62-200,Milano:2013sza}.
They are summarised in Table~\ref{t:much}. 
\begin{table}[b]
\caption{Values of temperature and chemical potentials at the chemical freeze-out from which we start the evolution of the chemical potentials.}
\label{t:much}
\begin{center}
\begin{tabular}{|c|c|c|c|}
\hline
$\sqrt{s_{NN}}$ [GeV] & $\tch$ [MeV] & $\mu_B$ [MeV] & $\mu_S$ [MeV] \\
\hline
7.7 & 144.3 & 389.2 & 89.5 \\
11.5 & 149.4 & 287.3 & 64.4 \\
19.6 & 153.9  & 187.9 & 45.3 \\
27 & 155.0 & 144.4 & 33.5 \\
39 & 156.4 & 103.2 & 24.5 \\
62.4 & 160.3 & 69.8 &16.7 \\
130 & 154.0 & 29.0 & 2.4 \\
200 & 164.3 & 28.4 & 5.6 \\
2760 & 156.0 & 0.0 & 0.0 \\
\hline
\end{tabular}
\end{center}
\end{table}
These values were found to reproduce the \emph{ratios} of hadron multiplicities and we keep them in order to satisfy that observation. 
Note that shifting these values slightly does not cause a big change in the \emph{shape} of the transverse momentum spectra.

The different sets of $(\tch,\mu_B,\mu_S)$ at the chemical freeze-out lead to different $T$-dependence of the chemical potentials
at different collision energies. As the energy is lowered, $\tch$ decreases. On the other hand, $\mu_B$ increases, since the ratio 
of baryons to antibaryons increases. The pion chemical potential always vanishes in chemical equilibrium at $\tch$. As a result of 
this behaviour, the ordering of curves for $\mu_\pi$ and $\mu_p$  is reversed: at the same temperature, $\mu_\pi$ for 
$\sqrt{s_{NN}} = 7.7$~GeV is the smallest, because it is evolved from the lowest $\tch$ of all, with the starting value $\mu_\pi(\tch) = 0$.
In contrast to that, $\mu_p$ for $\sqrt{s_{NN}} = 7.7$~GeV is the highest, because it is evolved from the highest $\mu_B$ at the 
chemical freeze-out. The ordering of $\mu_K$ follows that of $\mu_p$, because its starting value $\mu_S(\tch)$ follows that of $\mu_B$\footnote{%
A non-vanishing $\mu_B$ leads to an imbalance between strange baryons and antibaryons, most notably $\Lambda$ and $\bar \Lambda$. 
This spoils strangeness neutrality. In order to regain it, a non-zero $\mu_S$ is introduced, which then prefers $K^+$ over $K^-$ and restores
strangeness neutrality.
}.

The calculated chemical potentials as functions of temperature were tabulated and read in to DRAGON which used them in the simulations.


\section{Data and method}
\label{s:dam}

We fitted the transverse momentum spectra of (anti)protons, pions and kaons from the most central heavy-ion collisions measured by STAR and ALICE collaborations:
\begin{itemize}
\item Au+Au at  $\sqrt{s_{NN}} = 7.7, 11.5, 19.6, 27,$ and $39$~GeV \cite{STAR_7-39},
\item Au+Au at  $\sqrt{s_{NN}} = 62.4, 130, 200$~GeV \cite{STAR_62-200},
\item Pb+Pb at  $\sqrt{s_{NN}} = 2760$~GeV \cite{ALICE_piKp}. 
\end{itemize}
For $\sqrt{s_{NN}} = 130$~GeV the most central collisions are defined as 0-6\%, for all other remaining energies as 0-5\% of the total cross section.

The fitted intervals vary with energy and particle species. The upper value of $p_t$ was set to $2$~GeV for $\sqrt{s_{NN}} $ in the range 
$ 7.7 - 39$~GeV (for these energies this was also the upper end of the measured $p_t$ range) and for ALICE spectra, in order to remain in the region sensitive to collective effects and free of hard scattering processes. For $\sqrt{s_{NN}}$ in the range $ 62.4 - 200$~GeV, the published $p_t$ spectra reach only to about $0.7 - 1.1$~GeV, depending on the species. 
 
The lower end of the fitted intervals coincides with the lower end of the measured $p_t$ range, ranging from $p_t = 0.1$~GeV to $p_t = 0.4$~GeV depending on  the collision energy and the species.
The only exception (to be discussed in the next section) are the ALICE pion $p_t$ spectra - the data start at $p_t = 100$~MeV while the fit starts at $p_t = 250$~MeV.

Error bars on the data included statistical and systematic uncertainties as reported in the original papers \cite{STAR_7-39,STAR_62-200,ALICE_piKp}. We added the two kinds of uncertainties in quadrature.

For data fitting we have used the MADAI statistical analysis package \cite{madai2}.
Theoretical spectra were generated with the DRAGON package described in the previous section. 
The number of generated events was set to a value which guaranteed that the Monte Carlo statistical error was smaller than one third of the combined experimental error in the given $p_t$ bin for every bin and every particle species. This number was thus actually  set by antiprotons. The DRAGON spectra were generated typically in 400 training points in the three-parameter space (freeze-out temperature $\tk$, transverse flow gradient $\eta_f$, profile of the transverse velocity $n$). The typical run-time for one training point is 30 min to 5 hours depending on the number of events. Then the spectra were normalised to match the normalisation of the measured spectra, species by species, i.e. six independent normalisations.

The essence of MADAI is the Markov Chain Monte Carlo exploration of the parameter space weighted by the posterior probability, or likelihood for a particular point $\tk, \eta_f, n$ to represent the correct parameters given the experimental observations. The likelihood for points other than training points is not calculated from DRAGON (which is time consuming) but rather estimated from the Gaussian-Process based emulator trained on the training points. The output of MADAI is the best fit point in our three-parameter space. The uncertainties and correlations among parameters can be displayed as two-dimensional projections of the posterior distribution. As a final step (and a cross-check of MADAI) we calculated the $\chi^2$ value at the best fit point.


\section{Results} 
\label{s:res}

\begin{figure}
        \centering
	\includegraphics[width=0.49\textwidth]{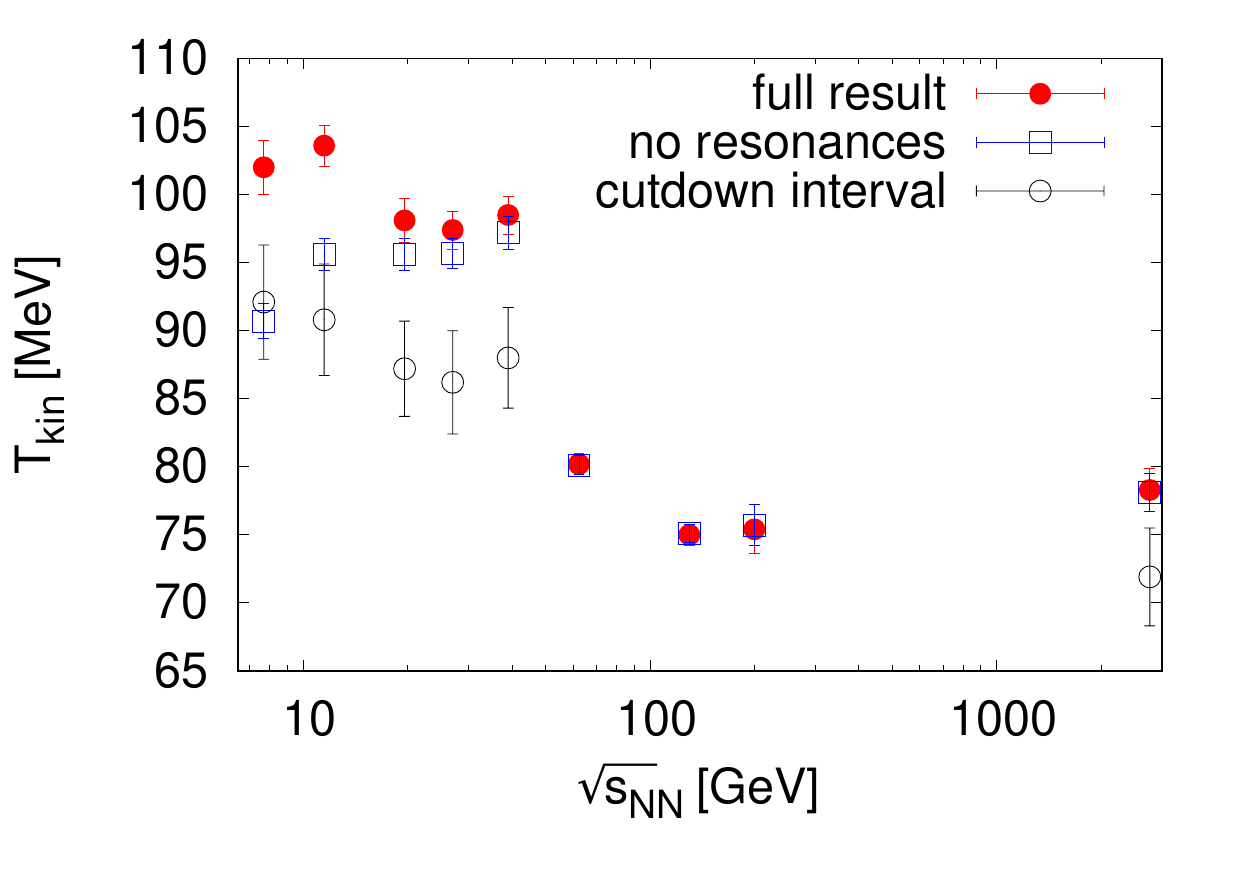}
	\includegraphics[width=0.49\textwidth]{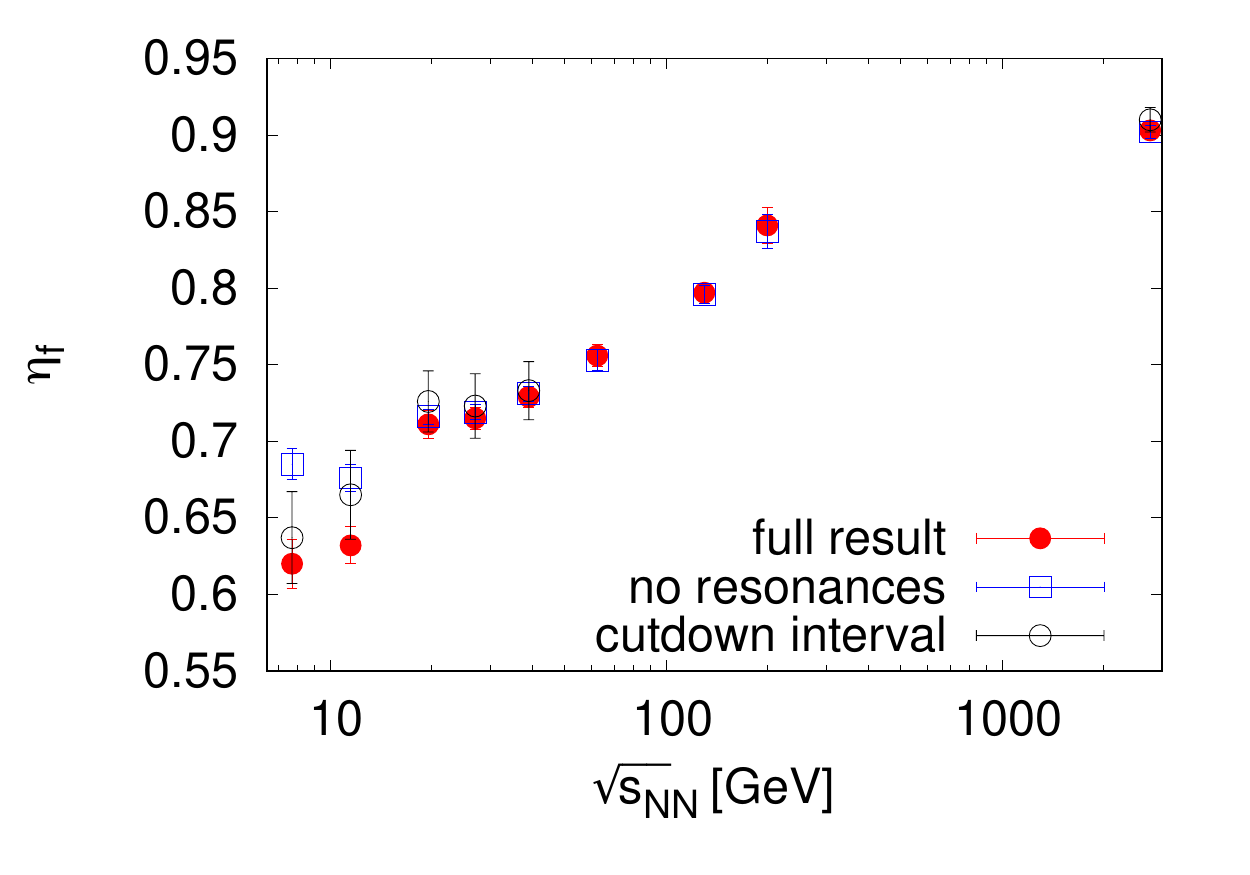}
	\includegraphics[width=0.49\textwidth]{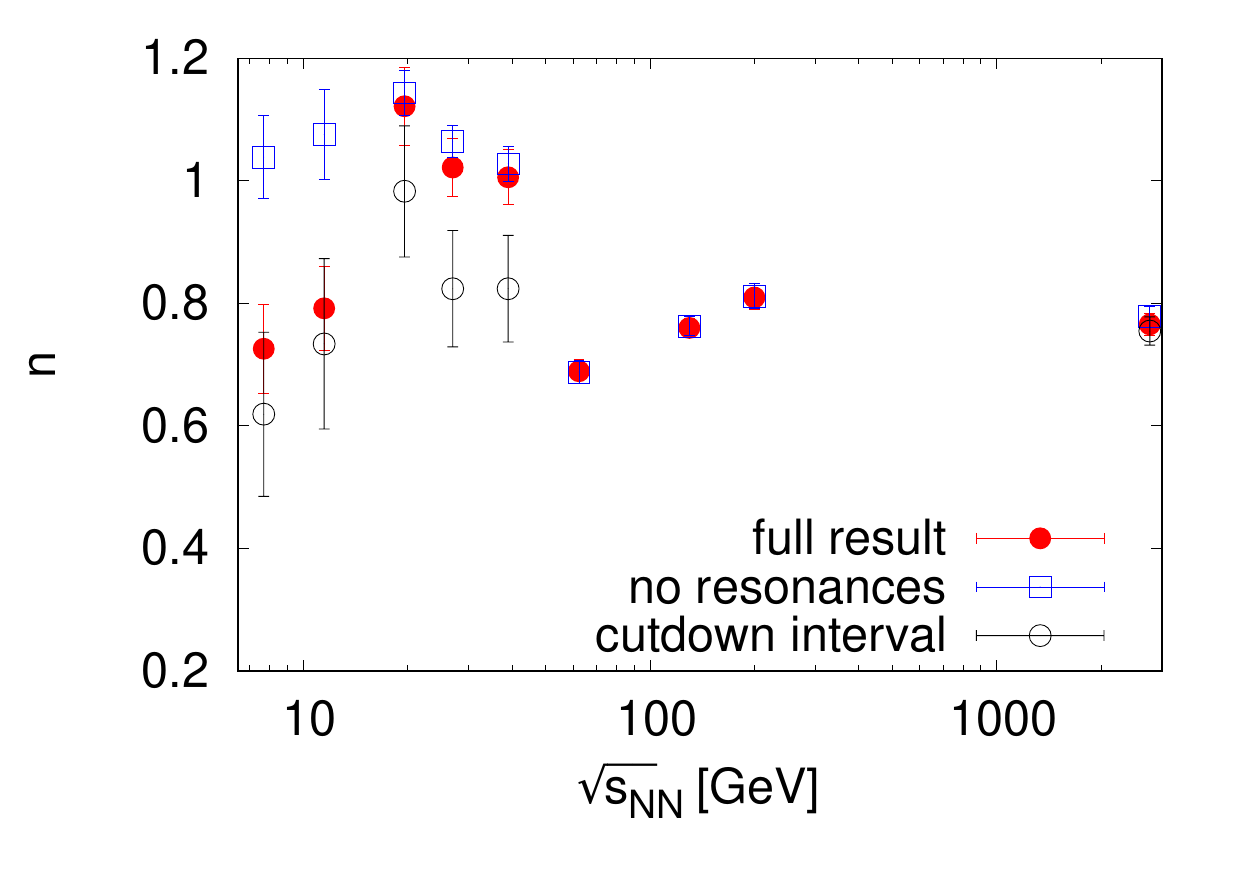}
	\includegraphics[width=0.49\textwidth]{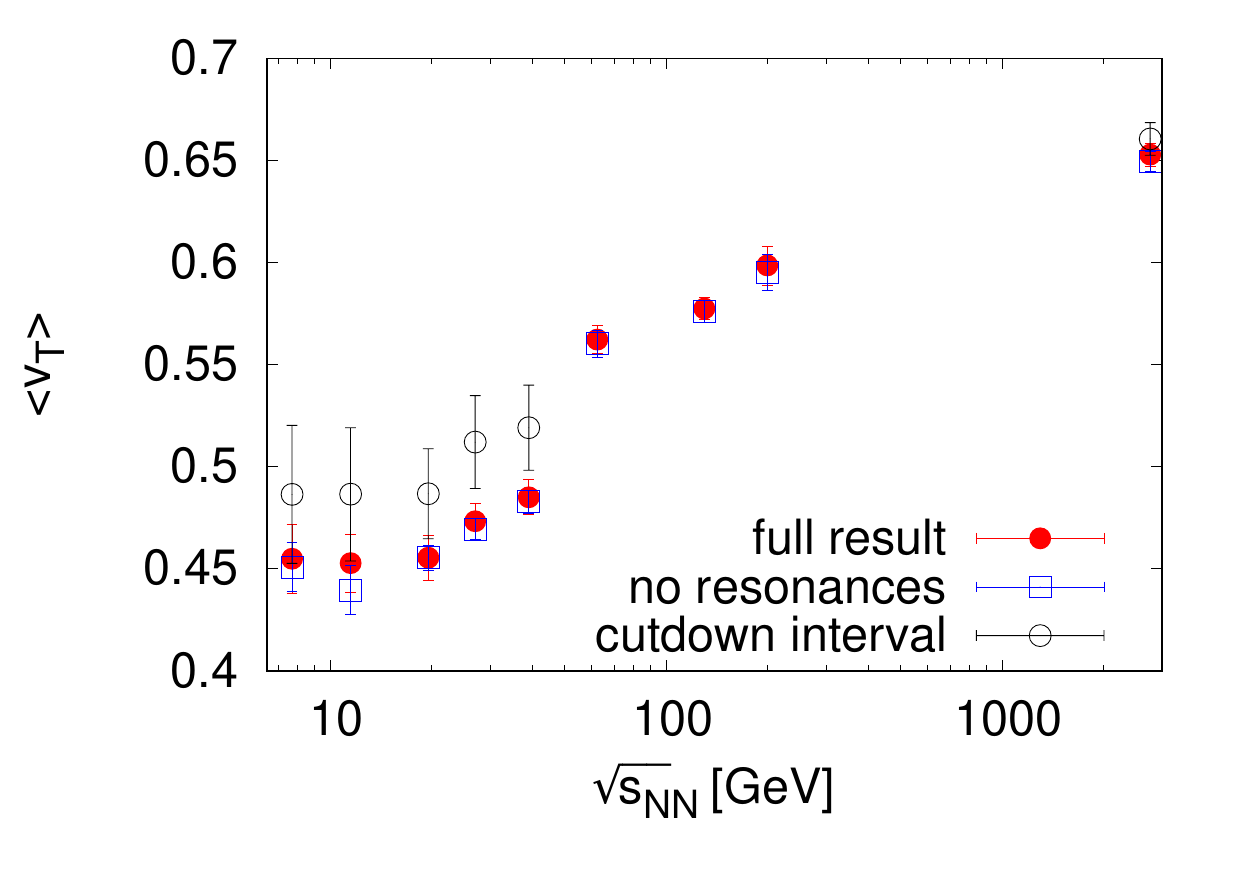}
\caption{\label{MainThing} Energy dependence of the freeze-out parameters: temperature (upper left), transverse flow gradient (upper right),
exponent of the transverse velocity profile (lower left), mean transverse velocity (lower right).}
\end{figure}

The main results of this paper, the energy dependences of the kinetic freeze-out temperature $\tk$, the transverse flow gradient $\eta_f$, the profile of the transverse velocity $n$, and the mean transverse velocity $v_t = 2 \eta_f/(n+2)$, are shown in Fig.~\ref{MainThing}. Three sets of results are plotted: the full results obtained from fits to $p_t$ spectra up to 2 GeV which include the contribution from the resonance decays (solid red circles), results from fits to $p_t$ spectra up to 2 GeV without the resonance contribution (blue squares) and finally results from fits to short (cut down) $p_t$ spectra up to $\sim 1$ GeV which match shorter $p_t$ intervals for $\sqrt{s_{NN}} = 62, 130, 200$ GeV with the resonance contribution included (empty black circles). For the full results we summarise the corresponding numerical values of $\tk, \eta_f$ and $n$ together with the $\chi^2/n_{dof}$ value  in Table~\ref{t:params}.

\begin{table}[bh]
\caption{\label{t:params}
Parameters of the best fits for different energies. These parameters were used also for the calculation of the resonance composition of the transverse momentum spectra. 
}
\begin{center}
\begin{tabular}{|c|cccc|}
\hline
$\sqrt{s_{NN}}$ [GeV]&$\tk$ [MeV]&$\eta_f$&$n$& $\chi^2/n_{dof}$\\ 
\hline
7.7 & $102.0 \pm 2.0$ &	$0.620 \pm 0.016$ &  $0.726 \pm 0.073$ & $0.83$ \\
11.6 & $103.6 \pm 1.5$  &	 $0.632 \pm 0.012$ & $0.792 \pm 0.069$ & $0.66$ \\ 
19 & $98.1 \pm 1.6 $ & 	$0.711 \pm 0.009$ &	$1.122 \pm 0.064$ & $0.38$ \\
27 & $97.4 \pm 1.4$ & $0.715 \pm 0.007$ & $1.022 \pm 0.048$ & $0.68$\\
39 & $98.5 \pm 1.4$ & $0.729 \pm 0.007$ & $1.006 \pm 0.045$ & $0.47$ \\
62 & $80.2 \pm 0.8$	& $0.756 \pm 0.007$&$0.689 \pm 0.020$ & $0.93$\\
130 & $75.0 \pm 0.8$ & $0.797 \pm 0.006$ & $0.760 \pm 0.015$ & $1.07$\\
200 & $75.4 \pm 1.8$ & $0.841 \pm 0.012$ & $0.810 \pm 0.020$ & $0.25$\\
2760 & $78.3 \pm 1.6$ & $0.903 \pm 0.005$ & $0.766 \pm 0.018$ & $0.32$\\
\hline
\end{tabular}
\end{center}
\end{table}

The freeze-out temperature decreases from $\tk \sim 100 - 104$ MeV at the lowest energies to $\tk \sim 75 - 80$ MeV at the highest energies. There is a little jump between $\sqrt{s_{NN}} = 39$ and $62.4$~GeV (full results, solid red circles).
This is most likely caused by the difference in $p_t$ intervals which are covered by the data. Recall that we usually fit spectra up to $p_t  = 2$~GeV, while 
the upper limit of the published  spectra at 64.2--200~GeV is around 0.8 GeV for mesons and 1.1~GeV for protons.
%
Indeed, once the $p_t$ intervals are matched (empty black circles), the jump is suppressed, although a steady decrease of $\tk$ with increasing $\sqrt{s_{NN}}$ remains. 
The effect of resonances is negligible at high energies while at the lowest energies they induce an upward shift in the temperature of the order of $10$~MeV. As we will see in the next section, this is caused by the lower population of particles from resonance decays  and also by the uniform relative contribution to $p_t$ spectra at high energies. 

The transverse flow gradient increases from $\eta_f = 0.62$ at $\sqrt{s_{NN}} = 7.7$~GeV to $\eta_f = 0.90$ at $\sqrt{s_{NN}} = 2760$~GeV. The effect of resonances is again significant only at the two lowest energies. The profile of the transverse velocity $n$ is the least sensitive parameter but one might conclude that it is roughly consistent with a constant value $n \sim 0.75$ once the lower energies are cut down to the short $p_t$ range.
The mean transverse velocity increases with energy from $v_t = 0.45$ to $v_t = 0.65$.

\begin{figure}
        \centering
	    \subfloat[\small $\sqrt{s_{NN}} = 7.7$ GeV]{\includegraphics[width=1.0\textwidth]{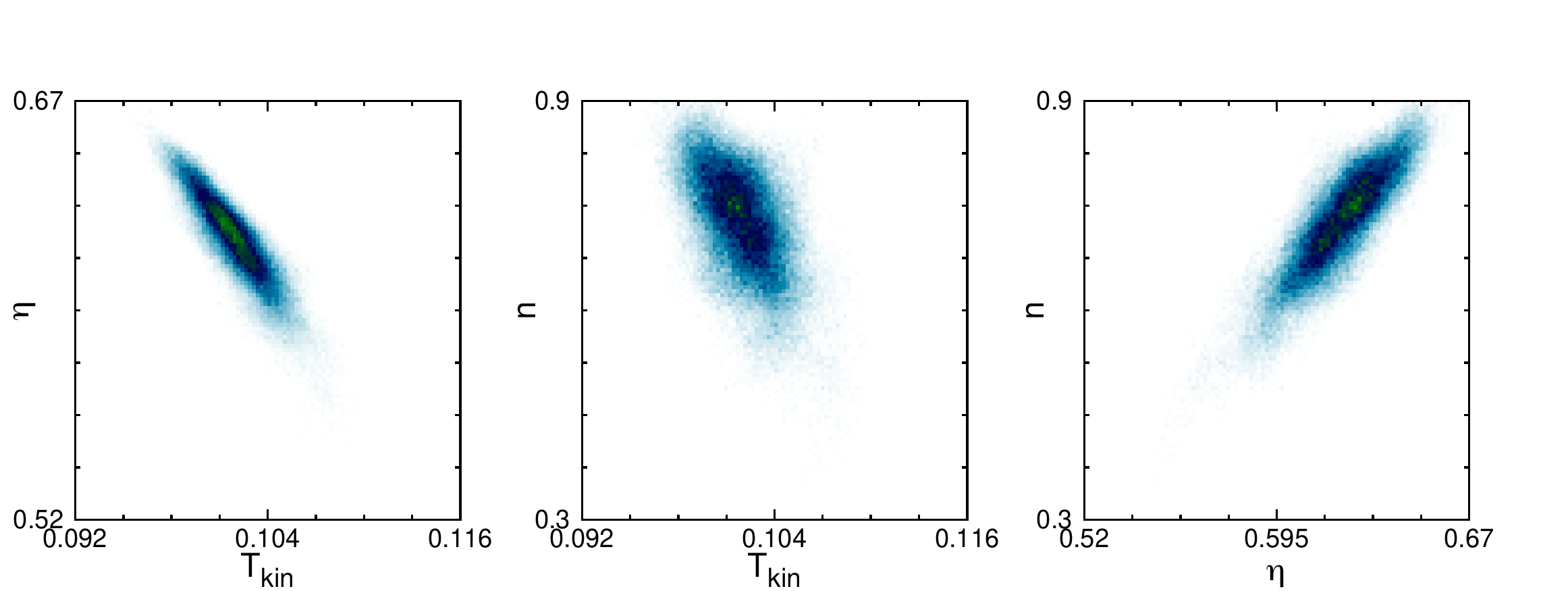}} \\
	    \subfloat[\small $\sqrt{s_{NN}} = 2760$ GeV]{\includegraphics[width=1.0\textwidth]{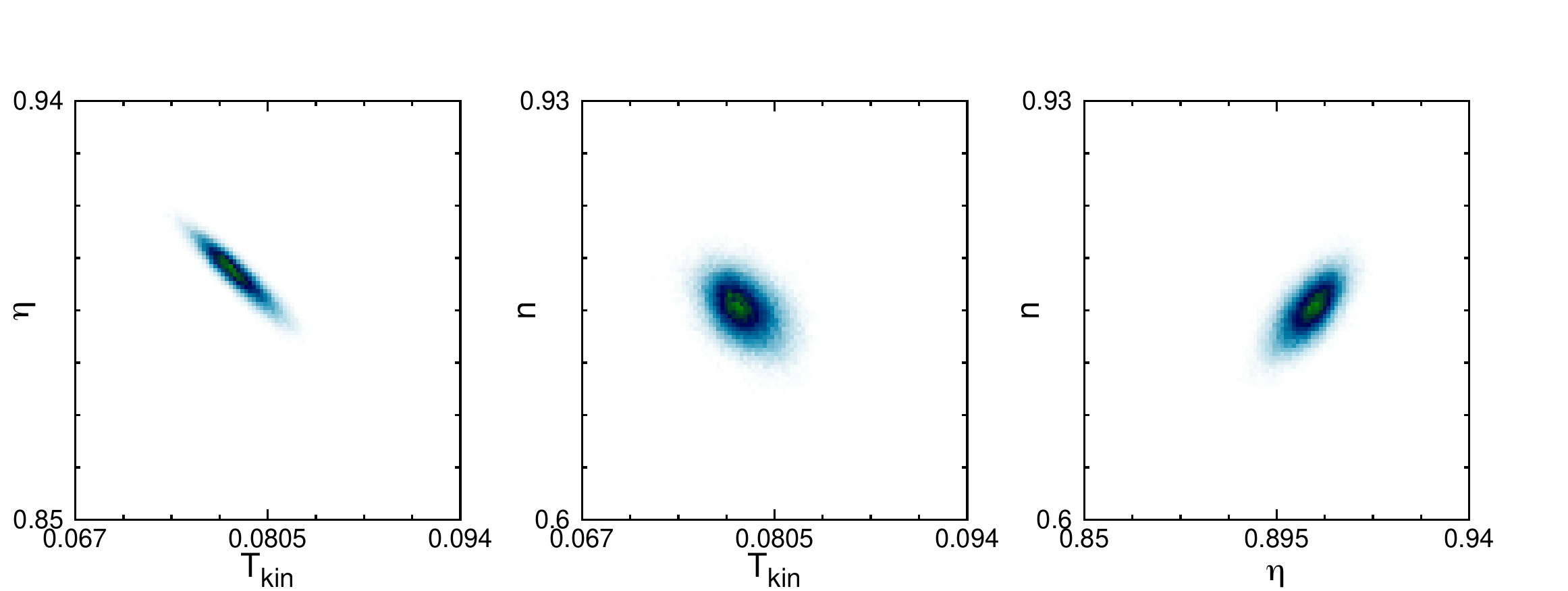}} 
\caption{\label{uncertainty_region} Two-dimensional projections of the posterior distribution for a) $\sqrt{s_{NN}} = 7.7$~GeV, b) $\sqrt{s_{NN}} = 2760$~GeV: $\eta_f - \tk$ (left column), $n - \tk$ (middle column), $n - \eta_f$ (right column).}
\end{figure}

In Fig.~\ref{uncertainty_region} two-dimensional projections of the posterior
distribution are plotted for a) $\sqrt{s_{NN}} = 7.7$ GeV and b) $\sqrt{s_{NN}} = 2760$ GeV. We can see the uncertainty region around the best fit parameters and also the correlation/anticorrelation among the parameters. For example, the temperature and the transverse flow gradient are anticorrelated in the sense that a slight decrease of $\tk$ can be compensated by an increase of $\eta_f$ with little effect on the quality of the fit. Also illustrated is a larger uncertainty of the best fit for $\sqrt{s_{NN}} = 7.7$ GeV, driven by the experimental errors (this is true also for $\sqrt{s_{NN}}$ in the range $ 11.5 - 39$ GeV).

\begin{figure}
\centering
\includegraphics[width=0.6\textwidth]{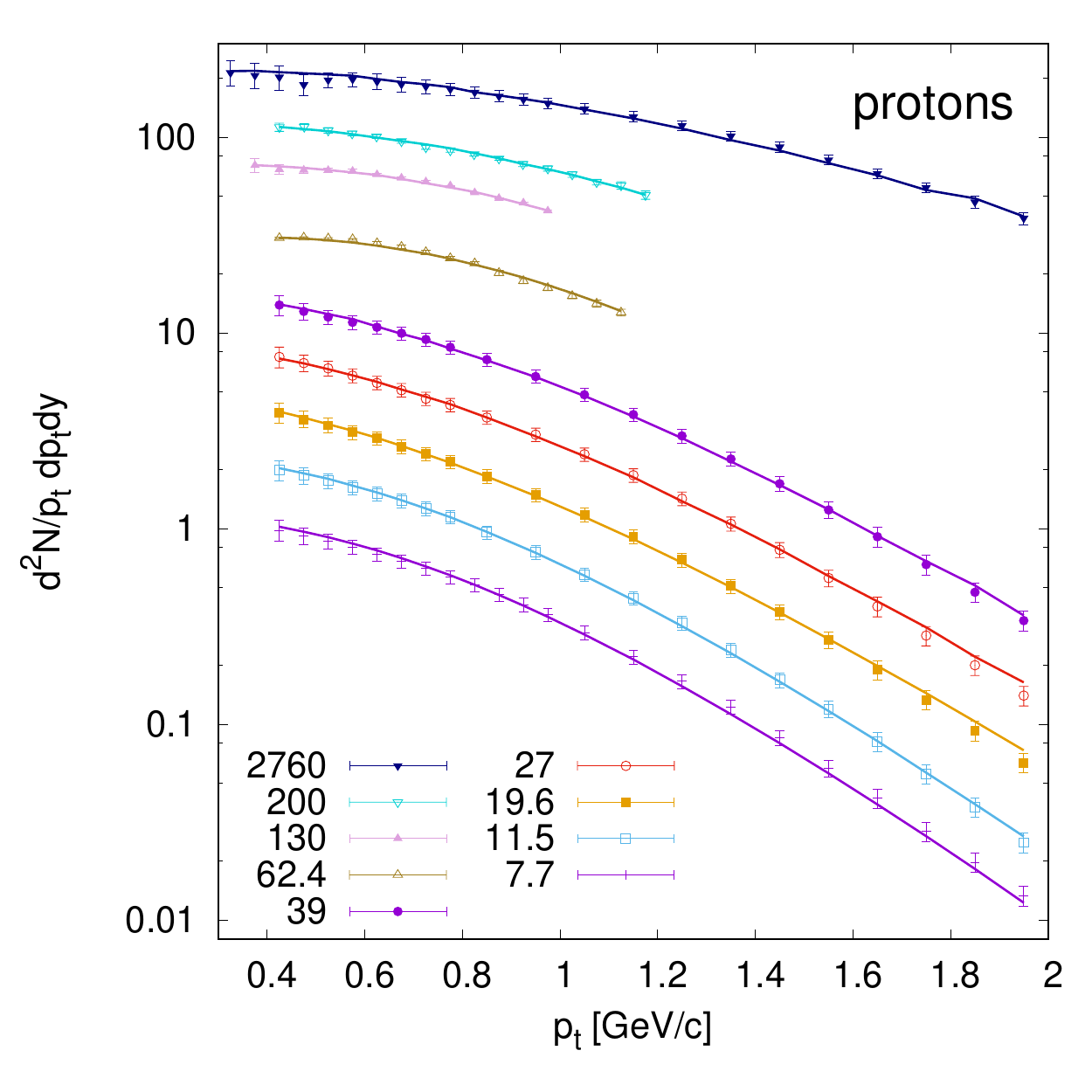}
\caption{\label{spectra} Transverse momentum spectra of protons for different energies. In order to display all spectra in one figure we divide data for $\sqrt{s_{NN}} = 7.7, 11.5, 19.6, 27, 39, 62.4, 130, 200, 2760$ GeV by factors $256, 128, 64, 32, 16, 8, 4, 2$ and $0.5$, respectively.}
\end{figure}

\begin{figure}
\centering
\includegraphics[width=0.6\textwidth]{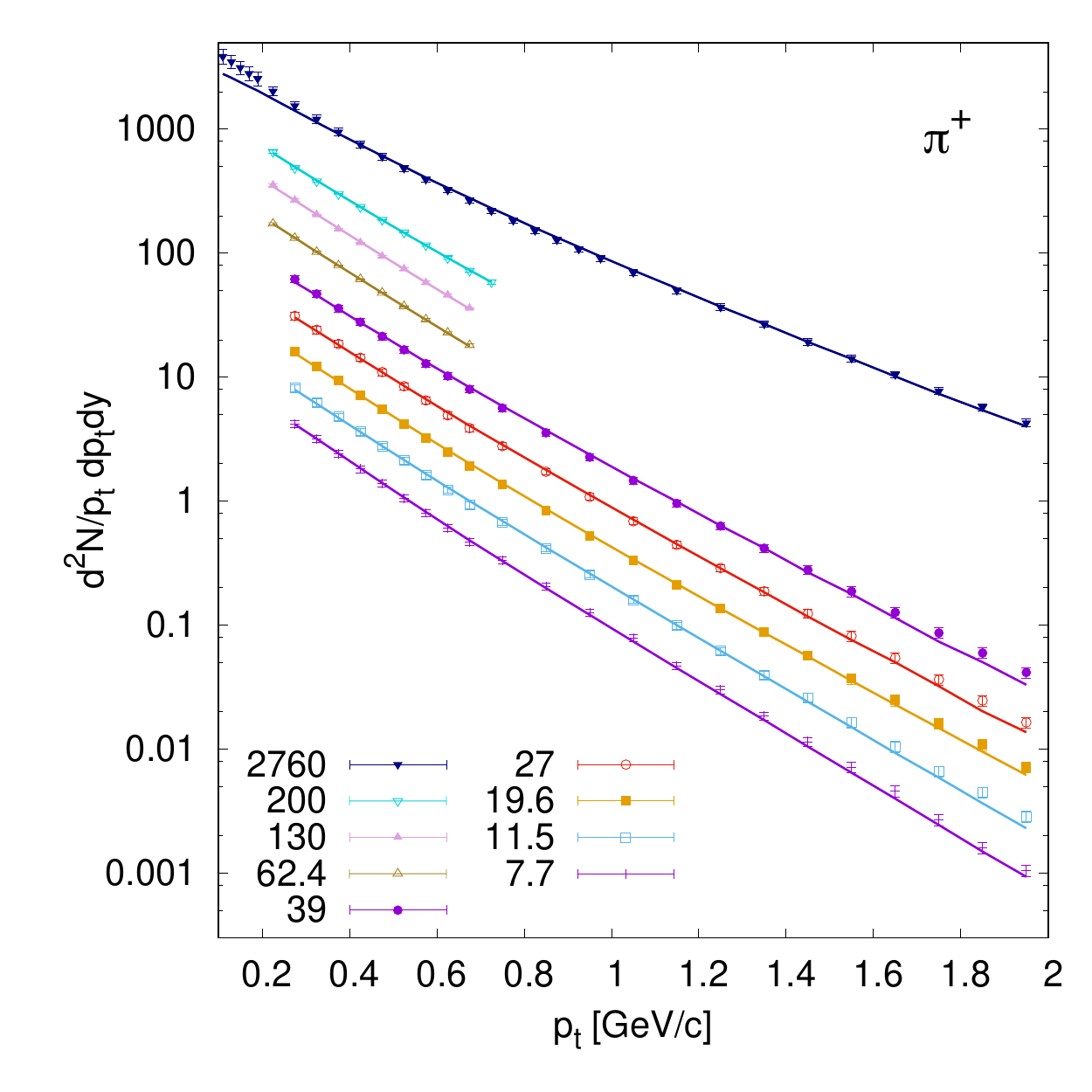}
\caption{\label{spectra1}Transverse momentum spectra of $\pi^+$ for different energies. In order to display all spectra in one figure we divide data for $\sqrt{s_{NN}} = 7.7, 11.5, 19.6, 27, 39, 62.4, 130, 200, 2760$ GeV by factors $256, 128, 64, 32, 16, 8, 4, 2$ and $0.5$, respectively. }
\end{figure}

\begin{figure}
\centering
\includegraphics[width=0.6\textwidth]{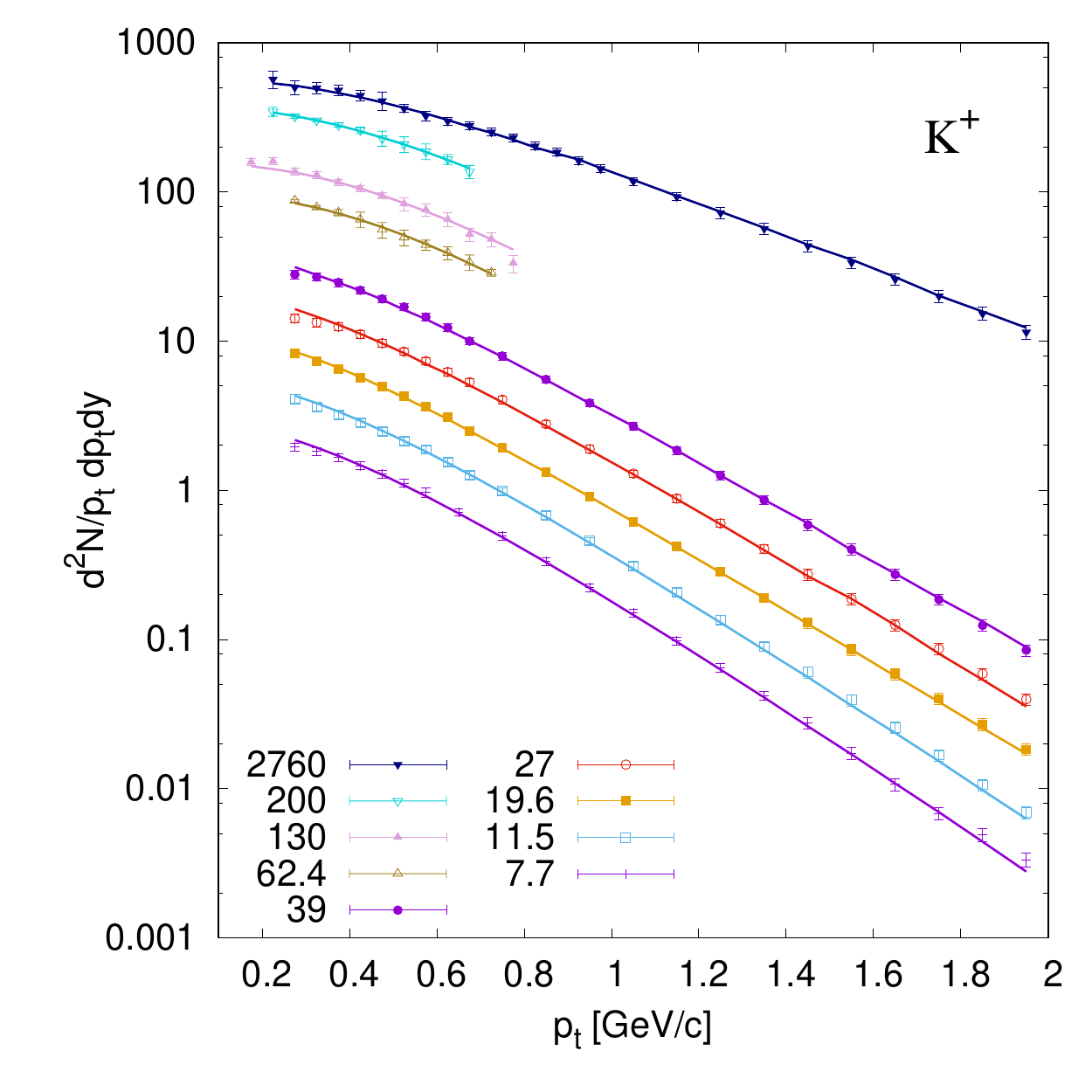}
\caption{\label{spectra2} Transverse momentum spectra of $K^+$ for different energies. In order to display all spectra in one figure we divide data for $\sqrt{s_{NN}} = 7.7, 11.5, 19.6, 27, 39, 62.4, 130, 200, 2760$ GeV by factors $256, 128, 64, 32, 16, 8, 4, 2$ and $0.5$, respectively.}
\end{figure}

\begin{figure}
\centering
\includegraphics[width=0.5\textwidth]{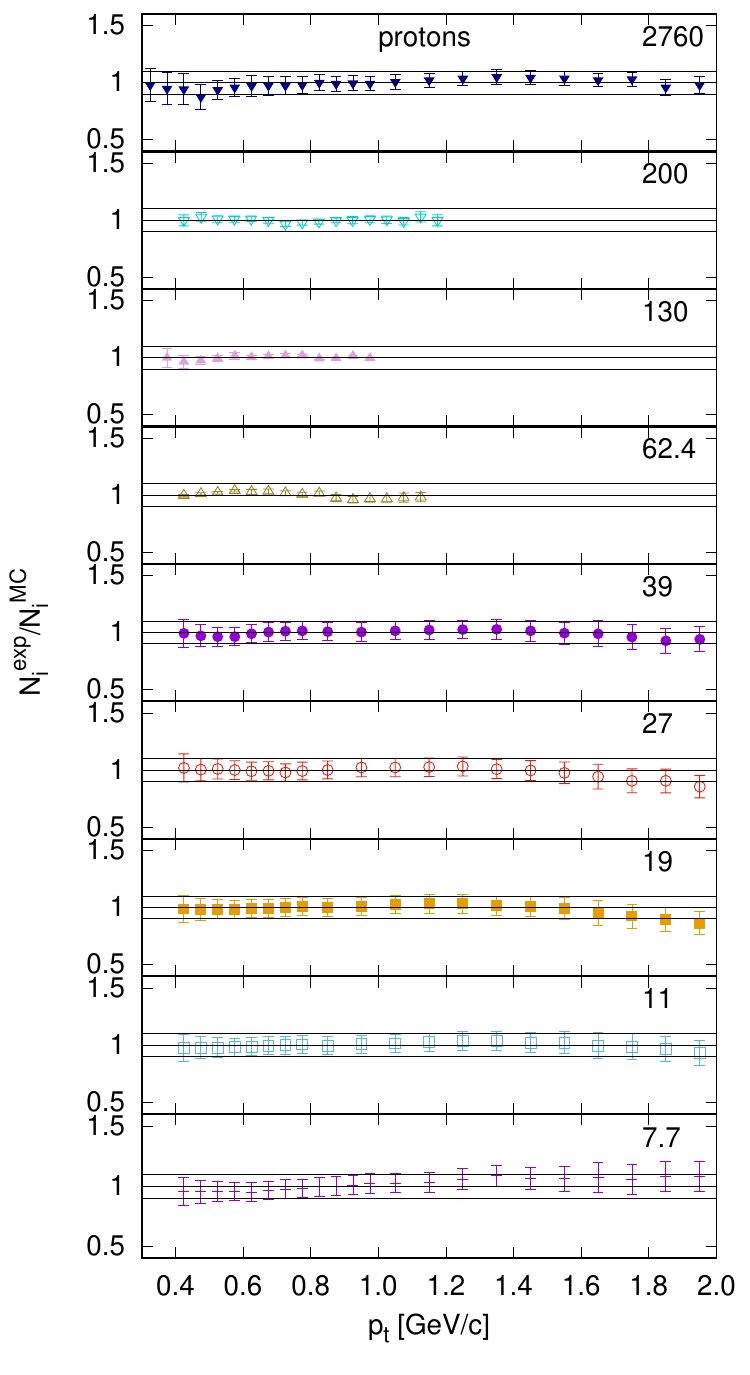}
\caption{\label{f:ratiosp} The ratio of data to Monte Carlo simulation, $N_i^{exp}/N_i^{MC}$ for $p_t$ spectra of protons. Different panels show the ratios for different collision energies. Horizontal lines indicate a band between 0.9 and 1.1.}
\end{figure}

\begin{figure}
\centering
\includegraphics[width=0.5\textwidth]{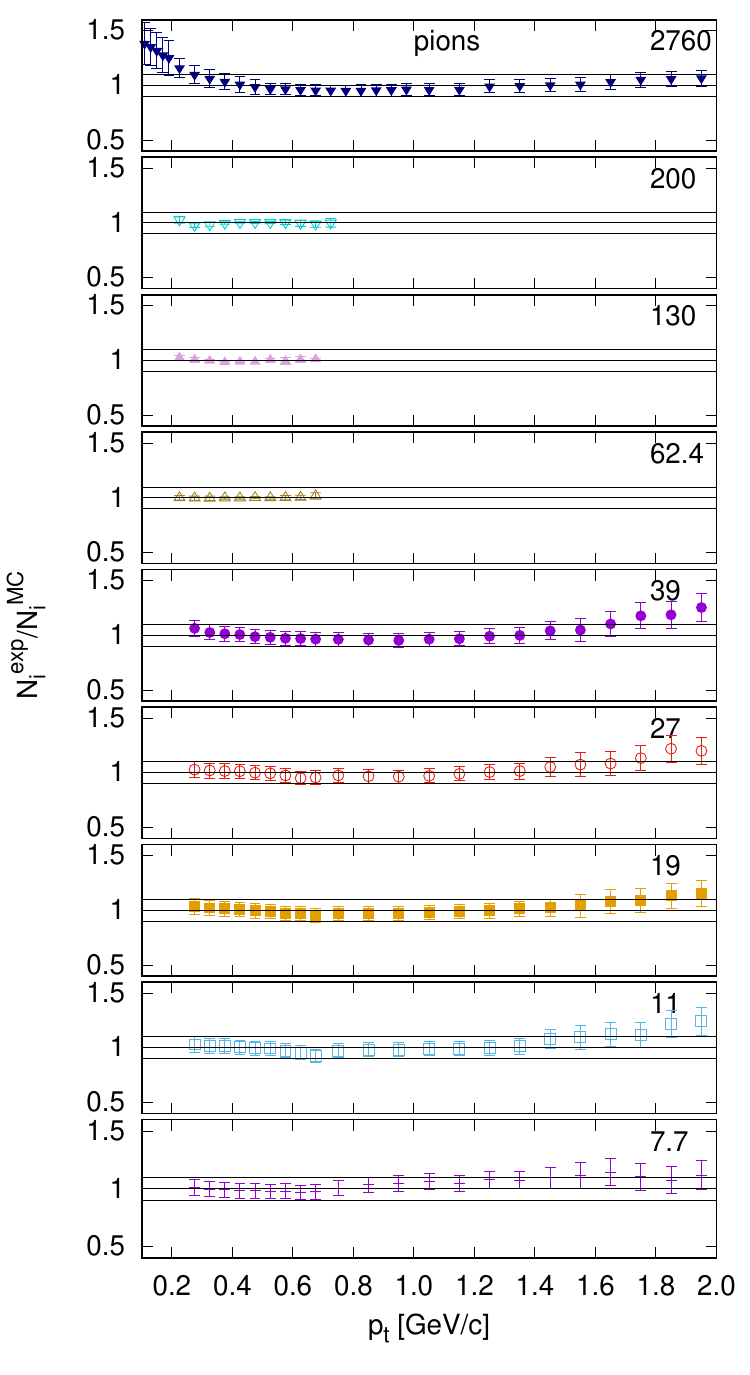}
\caption{\label{f:ratios} The ratio of data to Monte Carlo simulation, $N_i^{exp}/N_i^{MC}$ for $p_t$ spectra of $\pi^+$. Different panels show the ratios for different collision energies. Horizontal lines indicate a band between 0.9 and 1.1.}
\end{figure}

\begin{figure}
\centering
\includegraphics[width=0.5\textwidth]{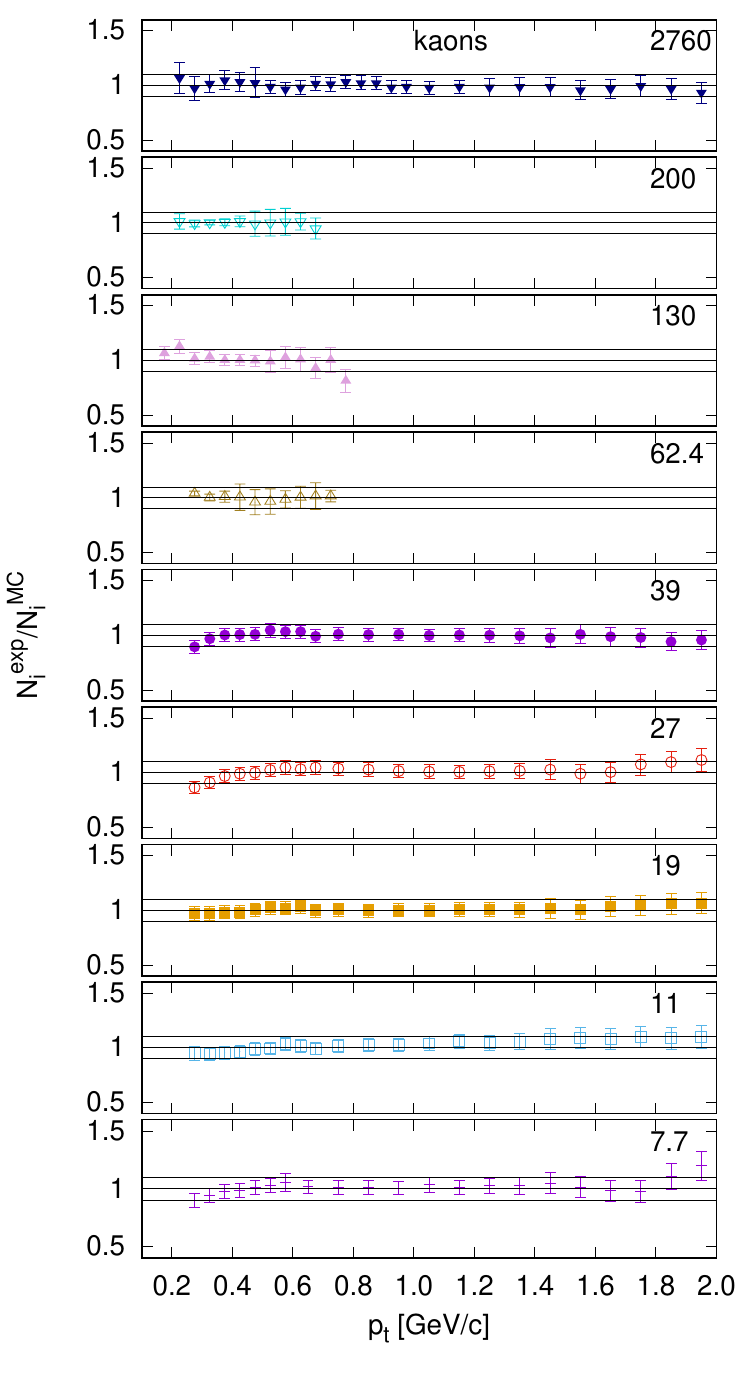}
\caption{\label{f:ratiosK} The ratio of data to Monte Carlo simulation, $N_i^{exp}/N_i^{MC}$ for $p_t$ spectra of $K^+$. Different panels show the ratios for different collision energies. Horizontal lines indicate a band between 0.9 and 1.1.}
\end{figure}

The simulated transverse momentum spectra of $p, \pi^+$ and $K^+$, corresponding to the best fit parameters (full results in Fig. \ref{MainThing} and Table~\ref{t:params}) are displayed in Figs. \ref{spectra}, \ref{spectra1}, \ref{spectra2} for different energies as solid lines along with the measured data points (including uncertainties) from STAR and ALICE.
The ratio of data to Monte Carlo simulation, $N_i^{exp}/N_i^{MC}$, are shown with errors in Figs.~\ref{f:ratiosp},~\ref{f:ratios}, and \ref{f:ratiosK}.

The general quality of the fits for all energies, particle species and the whole $p_t$ range is violated by the low $p_t$ pions at $\sqrt{s_{NN}} = 2760$~GeV where Monte Carlo underestimates the data (the first six $p_t$ bins were excluded from the fit in this case). 
Similar disagreement has been observed also in \cite{Li:2018jnm}.
There are at least two effects, both included in our simulation, which help populate this $p_t$ region: pions originating from the resonance decays and the nonzero pion chemical potential $\sim$100~MeV. We note that this disagreement with the data can be explained within the context of chemical non-equilibrium version of the statistical hadronization model \cite{Begun:2013nga}.  
Note, however,  that the enhancement of the pion spectrum at low $p_t$ with respect to the theoretical curve may well be caused by a specific shape of the freeze-out hypersurface in which the matter at larger distance from the longitudinal axis of the fireball freezes-out later than the matter in the middle\footnote{%
Technically, this is caused by a decreased flux of particles with higher $p_t$ through such a freeze-out hypersurface. Hence,
it is the high $p_t$ which becomes suppressed, thus leaving the non-suppressed low $p_t$ look like as it was enhanced. 
}. 
Such a hypersurface was used in the fits reported in \cite{Prorok:2015vxa}, which show better agreement with pions at low $p_t$. 


\section{Anatomy of the spectra}
\label{s:anat}

\begin{figure}
\centerline{%
\includegraphics[width=0.8\textwidth]{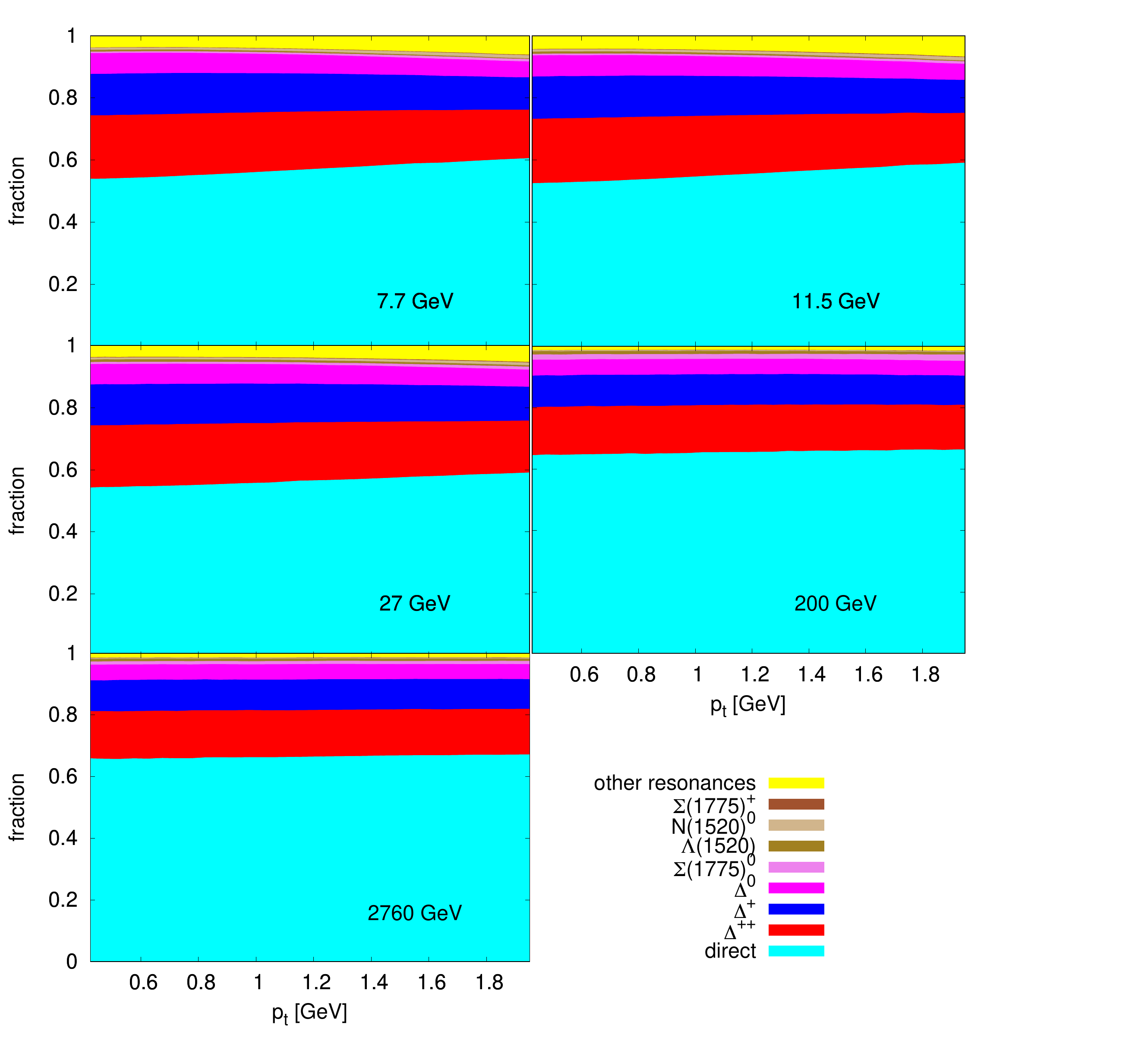}}
\caption{\label{f:a7.7}
The anatomy of the $p_t$ spectra of protons from collisions at $\sqrt{s_{NN}} = 7.7, 11.5, 27, 200$ and $2760$~GeV. 
Plotted are the ratios of protons of a given origin (direct or from a particular resonance) to the total number of protons as function of $p_t$.
The bands from the bottom show the relative contributions from direct protons, protons from $\Delta^{++}$, $\Delta^+$, $\Delta^0$, \dots
}
\end{figure}

\begin{figure}
\centerline{%
\includegraphics[width=0.8\textwidth]{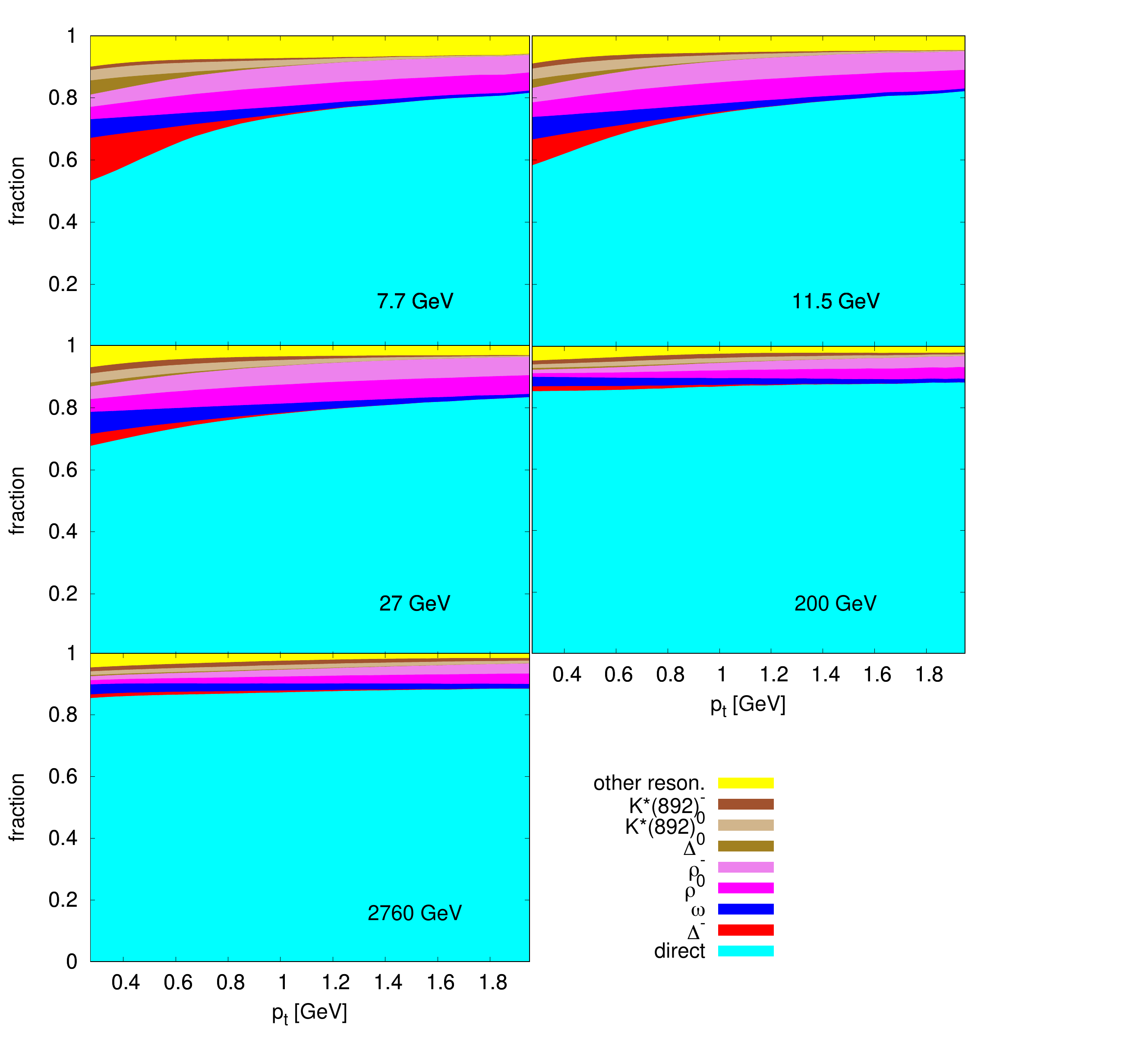}}
\caption{\label{f:a11}
The anatomy of the  $p_t$ spectra of $\pi^-$ from collisions at $\sqrt{s_{NN}} = 7.7, 11.5, 27, 200$ and $2760$~GeV. 
Plotted are the ratios of $\pi^-$ of a given origin (direct or from a particular resonance) to the total number of $\pi^-$ as functions of $p_t$.
The bands from bottom show the relative contributions from direct $\pi^-$, pions from $\Delta^{-}$, $\omega$, $\rho^0$, $\rho^-$, \dots
}
\end{figure}

\begin{figure}
\centerline{%
\includegraphics[width=0.8\textwidth]{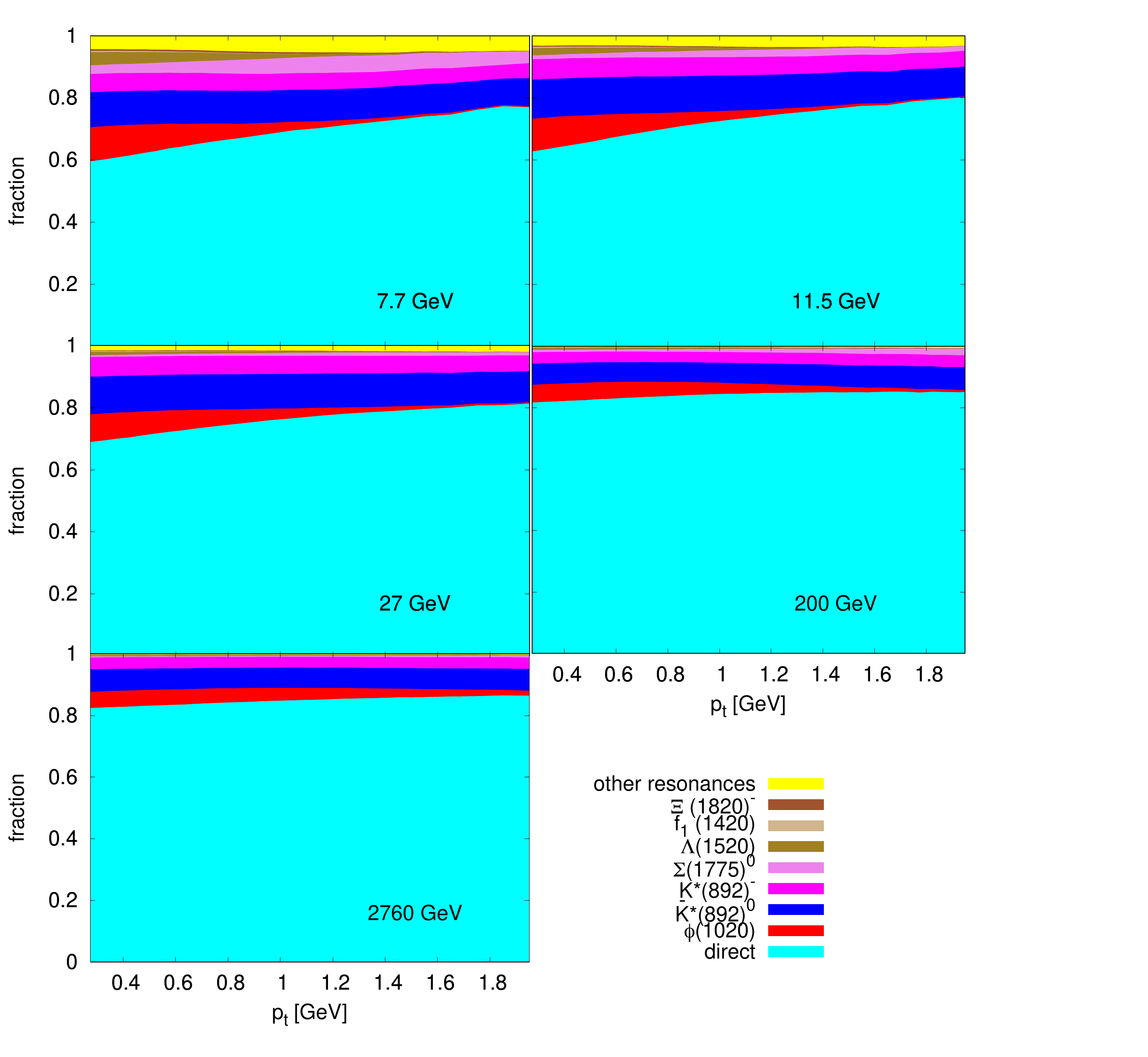}}
\caption{\label{f:a27}
The anatomy of the $p_t$ spectra of $K^-$ from collisions at $\sqrt{s_{NN}} = 7.7, 11.5, 27, 200$ and $2760$~GeV. 
Plotted are the ratios of $K^-$ of a given origin (direct or from a particular resonance) to the total number of $K^-$.
The bands from bottom show the relative contributions from direct $K^-$, kaons from $\phi$, ${\bar K}(892)^0$, $K(892)^-$, \dots
}
\end{figure}

Let us come back to the observation, that the effect of resonance decays on the extracted $\tk$ is negligible at high $\sqrt{s_{NN}}$, while 
they cause a slight upward shift at low $\sqrt{s_{NN}}$. In order to understand this we 
have looked at the composition of the $p_t$ spectra---how does the origin of all observed pions (kaons, protons) depend on $p_t$? Results for five chosen energies, $\sqrt{s_{NN}} = 7.7, 11.5, 27, 200, 2760$~GeV, are plotted in Figures~\ref{f:a7.7}, 
\ref{f:a11} and \ref{f:a27}. The values of the parameters used in the calculations of these figures are summarised in Table~\ref{t:params}.


We observe some systematics connected with the change of the collision energy:
\begin{enumerate}
\item The fraction of resonance-produced hadrons decreases as the collision energy goes up. This may seem surprising at the first sight.  However, this behaviour is closely connected with the scenario of partial chemical equilibrium. The share of resonance production is high at the moment of the chemical freeze-out. Afterwards, however, the temperature decreases and even though chemical potentials develop for all resonance species, the weight of particle production moves from resonances towards directly produced particles. 
\item At lower collision energies the resonances populate more the low $p_t$ interval while at 200~GeV and higher  the share of resonance production seems to be rather flat as function of $p_t$. This change is actually gradual. A closer inspection reveals that the feature is mainly brought in by the $\Delta$ resonance. 
Its decay happens closely above the threshold, so that daughter particles do not acquire high momentum. In combination with small transverse expansion velocity this causes that pions from such decays stay at low $p_t$. (A similar argument applies for kaons from the decay of $\phi$.)
Let us also stress that at lower energies the contribution from \emph{baryon} resonances (most importantly $\Delta$) to pion production is more
important than at higher energies. 
A semi-quantitative discussion of this feature is presented in \ref{s:hadmom}.
\end{enumerate}
We have also explored how these results change if $\tk$ is varied within the confidence regions as indicated in Fig.~\ref{uncertainty_region}.
The changes in the composition of the $p_t$ spectrum can be barely seen in the figures, so we refrain from showing them here. If $\tk$ is decreased
to the edge of the confidence interval (which is usually a shift smaller than 2~MeV), generally the fractions of directly produced hadrons increase their values by a few per cent.
The increase is slightly larger at low $p_t$ than at high $p_t$, and also it is larger at low $\sqrt{s_{NN}}$ than at high $\sqrt{s_{NN}}$.


\section{Conclusions}
\label{s:conc}

Our results clearly show that with the increase of the collision energy the fireball develops stronger transverse expansion and 
cools further down. The former feature is well expected. The  decrease  of temperature by more than 20~MeV can also be 
understood:  more energy and entropy is deposited in the collision, which expands to a larger volume. At larger volume the local (longitudinal) flow gradients
are smaller and so is the expansion rate. Thus, also the scattering rate at the freeze-out drops lower, and so does the temperature.
To confirm this scenario we would need to extract the sizes of the fireball with the help of femtoscopy. This goes beyond 
the scope of the present paper and we leave it for further investigation. 

Although the excitation function of the kinetic freeze-out temperature seems to show a sharp step between 39 and 62.4~GeV, it would be premature 
to make any conclusions out of this. The feature may be connected with the different coverage of $p_t$ intervals 
in the different data sets. 

It is interesting to observe that the results obtained with the full model with resonances coincide with those obtained with only 
directly produced hadrons, i.e. basically just with fitting the formula ~(\ref{e:dirspec}). This is seen for all but the two lowest collision 
energies. It is crucial here that the partial chemical equilibrium was assumed, as a consequence of the cooling of the
fireball between the chemical and the thermal freeze-out. The cooling is most pronounced at high collision energies where the temperature drops from 
about 160~MeV to some 80~MeV, i.e.~by 80~MeV. In contrast to that, at $\sqrt{s_{NN}} = 7.7$~GeV it went down only by roughly 
40~MeV between 144 and 102 MeV. In view of these numbers it is clearly understood that the influence of resonances becomes less important
at high energies. 

We hoped originally that  we would be able to fit the low $p_t$ enhancement of the pion spectra at the LHC, since pions develop
chemical potential about 90~MeV at the kinetic freeze-out temperature. Note that a successful fit was obtained with 
the non-equilibrium Cracow single freeze-out model\footnote{%
Note that in Cracow single freeze-out model, non-equilibrium refers to the feature that chemical equilibrium is not reached
even at the chemical freeze-out, unlike in our model here. 
} %
with a possible admixture of pion condensation \cite{Begun:2015ifa}.
Nevertheless, the enhancement may also be caused by a specific shape of the freeze-out hypersurface used in \cite{Begun:2015ifa} which corresponds 
to an inside-out freeze-out in transverse direction \cite{raduska}. 
Such features have never been explored in the blast-wave-like fits and this opens a question whether it is worth studying.

The least well determined parameter of the model is the exponent $n$. It might be consistent with a constant value 
if the cuts on spectra are applied.
Clearly, spectra in broader $p_t$ intervals also at $\sqrt{s_{NN}} = 62.4, 130$, and 200~GeV are necessary in order to settle this uncertainty.
Nevertheless, full experimental results seem to indicate that $n$ decreases as the $\sqrt{s_{NN}}$ 
grows. This may be purely kinematically determined feature. The transverse velocity profile of eq.~(\ref{e:vn}) is 
constructed so that it never grows above 1. 
However, in real fireball the velocity at the edge may get closer to this limit.
At this point, relativity kicks in. As the matter is locally boosted to a higher transverse velocity, 
then due to the Lorentz transformation of the velocity into the global frame the dependence $v_t(r)$ begins to level off. 
The concave dependence  in our model is parametrised by $n$ with values smaller than 1. 

We do not want to conclude without mentioning the results of \cite{Mazeliauskas:2019ifr} which appeared just days before the 
submission of our paper. Apparently, spectra from the ALICE experiment have been fitted there with the same model as we 
used here, but the resonance contribution has been determined with the help of a novel method \cite{Mazeliauskas:2018irt}
that avoids MC simulations. At $\sqrt{s_{NN}}= 2.76$~TeV their result in the partial chemical equilibrium in central Pb+Pb collisions
is $\tk=127\pm 2$~MeV. Presently, we do not have an explanation for this difference. To the extent we can judge from 
\cite{Mazeliauskas:2019ifr}, the only difference is, that the lists of stable species in the partial chemical equilibrium 
there and in our work are different.
It will be important and interesting to sort out this tension in the near future. 

In this paper we presented results of a pilot study where only data from \emph{central} collisions have been fitted. A more comprehensive study 
which will also include the centrality dependence is being elaborated and will be published later.


\ack
We gratefully acknowledge financial support by  
VEGA 1/0348/18 (Slovakia), 
by the Czech Science Foundation via grant No. 17-04505S, and by the Ministry of Education of the Slovak Republic via project FEPO.
Computing was performed in the High Performance Computing Center of the Matej Bel University in Bansk\'a Bystrica using the HPC infrastructure acquired in project ITMS 26230120002 and 26210120002 (Slovak infrastructure for high-performance computing) supported by the Research \& Development Operational Programme funded by the ERDF.


\appendix

\section{Estimates of hadron momentum from resonance decays}
\label{s:hadmom}

In this appendix we seek some back-of-the-envelope understanding of why the resonance contribution to the $p_t$-spectra is different 
at different energies. This definitely does not replace the full calculation which convolutes the spectra of resonances with the phase space 
available for the decay, and which was effectively done when we simulated the spectra and calculated the ratios shown in Figs.~\ref{f:a7.7}-\ref{f:a27}.
A closer inspection of those figures shows that contributions to the $p_t$ spectrum which are not flat in $p_t$ are due to resonances with 
masses close above the sum of the masses of the daughter particles. The most notable is the decay $\Delta \to N\pi$. In such decays,
the decay products acquire only small kinetic energy\footnote{%
The natural scale to determine what is small and what is big, is provided by the temperature. Energies, which are  comparable to $T$, we call small.
}.
For example, the total kinetic energy released in the decay of $\Delta$ is 155~MeV. 


Let us try to understand, why the contributions of resonance decays populate pion spectra mainly at low $p_t$ in collisions at 
$\sqrt{s_{NN}} = 7.7$~GeV, while they are flat in $p_t$ at 200~GeV and higher energies. 

At lower collision energies, also the transverse expansion of the fireball is weaker. The heavy resonances follow more closely the collective velocity of the fluid and do not depart from it with thermal velocities too much. Thus there is only small boost in the transverse direction and pions (nucleons) are produced with small momenta from the decays of such resonances. However, at higher collision energies, the transverse expansion is stronger and the heavy resonances also obtain a stronger boost in that direction, depending on their position in the fireball. Thanks to this boost the resonance decays can also populate daughter particles with higher $p_t$. 

We illustrate these qualitative considerations with some simple estimates. First,  the momentum of the pion from a decay of $\Delta$ in the rest frame of the $\Delta$ is
\begin{equation}
p = \frac{\sqrt{(M^2 - m_\pi^2 - m_N^2)^2 - 4m_\pi^2 m_N^2}}{2M}\,  ,
\end{equation}
where $M$ is the mass of the resonance. For the decay $\Delta \to N\pi$ we obtain $p = 227.7$~MeV. This gives the total energy for the pion 266.8~MeV and its share of the released kinetic energy is 127~MeV (out of the quoted 155~MeV). 

Now we want to estimate the maximum $p_t$ that we can expect for a pion emitted 
from a $\Delta$-resonance which is boosted transversely due to expansion of the fireball.

If the $\Delta$ resonance moves with velocity $v$, and the pion is emitted in the direction of the collective velocity, its transverse momentum is 
boosted to the lab frame 
\begin{equation}
p_{lab} = \gamma p + v\gamma E_\pi\,  ,\qquad \mathrm{with} \quad \gamma = \frac{1}{\sqrt{1-v^2}}\,  ,
\label{e:boost}
\end{equation}
where $E_\pi$ is the energy of the pion. 

In addition to the flow velocity, the resonance also has some random thermal velocity, which we also want to estimate. 
Since its total energy is the sum $m+E_{kin}$ and the kinetic energy is given by the temperature, we have
\begin{equation}
\gamma_{th} = \frac{m+E_{kin}}{m} = \frac{m+T}{m}
\end{equation}
which gives
\begin{equation}
v_{th} = \frac{\sqrt{2mT + T^2}}{m+T}\,  .
\end{equation}

Let us now estimate the typical $p_t$ scale which can be reached by pions from $\Delta$ decays. We do this for two values 
of $\sqrt{s_{NN}}$, the lowest and the highest  energy. 

\begin{description}
\item[7.7~GeV] The transverse expansion velocity at the edge of the fireball is 0.62, and the $\gamma$ factor is then 1.275.
Suppose a $\Delta$ resonance moving with this velocity. In order to estimate the maximum $p_t$ a pion can obtain from 
a $\Delta$-decay, suppose the pion from this $\Delta$ moves in the same direction. According to eq.~(\ref{e:boost}) it acquires the
$p_t$ of 500~MeV. On top of this the $\Delta$ also moves with random thermal velocity about 0.37 in the fluid rest frame. 
This velocity is directed randomly, so it will boost some pions to higher $p_t$ and some to lower $p_t$. We 
conclude that pions from $\Delta$ decays will reach up to $p_t$ of about 500~MeV.
\item[2760~GeV] We proceed similarly for the fireball at the highest energy. 
The transverse expansion velocity at the edge of the fireball is 0.903 and the $\gamma$ factor is then 2.327.
If the pion from the decay is boosted with this velocity according to eq.~(\ref{e:boost}), it acquires the momentum 1090.5~MeV.
Thermal velocity of the $\Delta$ is 0.33.
\end{description}

We basically see that there is not a big difference between the thermal boost velocities in the two cases. 
Hence, we expect that at 7.7~GeV the contribution to pion production from $\Delta$ decays will reach up to 
500~MeV, while at 2760~GeV it should go to 1090~MeV. These limits are then blurred by the thermal smearing, which 
is comparable in both cases.
This explains the different $p_t$ dependences of the relative contribution to pion $p_t$ spectra at different energies.




\end{document}